\documentclass[preprint,showpacs,titlepage,aps,prd,tightenlines,amsmath,byrevtex,nofootinbib]{revtex4}

\usepackage{graphicx}

\def\lsim{\raise0.3ex\hbox{$\;<$\kern-0.75em\raise-1.1ex
\hbox{$\sim\;$}}}
\def\gsim{\raise0.3ex\hbox{$\;>$\kern-0.75em\raise-1.1ex
\hbox{$\sim\;$}}}

\begin{document}


\title{Large-$\theta_{13}$ Perturbation Theory of Neutrino Oscillation for Long-Baseline Experiments}

\author{Katsuhiro Asano} 
\email{katsuhiro.asano@gmail.com}

\author{Hisakazu Minakata}
\email{minakata@tmu.ac.jp}

\affiliation{Department of Physics, Tokyo Metropolitan University \\
1-1 Minami-Osawa, Hachioji, Tokyo 192-0397, Japan}

\date{June 3, 2011}

\vglue 1.4cm

\begin{abstract}

The Cervera {\it et al.} formula, the best known approximate formula of neutrino 
oscillation probability for long-baseline experiments, can be regarded as a 
second-order perturbative formula with small expansion parameter 
$\epsilon \equiv \Delta m^2_{21} / \Delta m^2_{31} \simeq 0.03$ under the 
assumption $s_{13} \simeq \epsilon$. If $\theta_{13}$ is large, as 
suggested by a candidate $\nu_{e}$ event at T2K as well as the recent global analyses, 
higher order corrections of $s_{13}$ to the formula would be needed for better accuracy. 
We compute the corrections systematically by formulating a perturbative framework 
by taking $\theta_{13}$ as $s_{13} \sim \sqrt{ \epsilon } \simeq 0.18$, which 
guarantees its validity in a wide range of $\theta_{13}$ below the Chooz limit. 
We show on general ground that the correction terms must be of order $\epsilon^2$. 
Yet, they nicely fill the mismatch between the approximate and the exact formulas 
at low energies and relatively long baselines. 
General theorems are derived which serve for better understanding of 
$\delta$-dependence of the oscillation probability. 
Some interesting implications of the large $\theta_{13}$ hypothesis are discussed.

\end{abstract}


\maketitle

\section{Introduction}

One of the most important progresses in particle physics in the last decades is 
the discovery of neutrino masses \cite{kajita} and the lepton flavor mixing \cite{MNS}. 
It was done through observing neutrino oscillation phenomena and it constitutes, 
up until this moment, only available experimental method for measuring 
lepton mixing parameters. 
$\Delta m^2_{32}$ and $\theta_{23}$ are determined by atmospheric neutrino observation by Super-Kamiokande 
\cite{SKatm-evidence,SKatm-L/E,SKatm-parameter}, and then by accelerator 
neutrino experiments \cite{K2K-full,MINOS}. 
$\Delta m^2_{21}$ and $\theta_{12}$ are measured independently by two 
types of experiments, the KamLAND reactor experiment 
\cite{KamLAND-evidence,KamLAND-spetrum,KamLAND-13} 
and the solar neutrino observation using various experimental techniques. 
For the latest results and for a review of the solar neutrino experiments see e.g., 
\cite{SNO-lowE,SK-solar-III} and \cite{solar-review}, respectively.
The remaining mixing angle $\theta_{13}$ is being explored by the ongoing 
and the upcoming accelerator \cite{T2K,NOVA} and reactor neutrino experiments \cite{DCHOOZ,RENO,Daya-Bay}. 
If it turned out that $\theta_{13}$ is not too small, we may proceed to 
measure CP violation by the lepton Kobayashi-Maskawa (KM) \cite{KM} phase 
$\delta_{\ell}$, to which we refer just $\delta$ in this paper.

It is expected that precision measurement is required to determine $\delta$ 
because CP violation effect is tiny due to suppression by the two small factors, 
$\Delta m^2_{21} / \Delta m^2_{31}$ and the Jarlskog coefficient 
$J \equiv c_{12} s_{12} c_{23} s_{23} c^2_{13} s_{13}$ \cite{jarlskog}. 
Therefore, understanding of full complexity of neutrino oscillation phenomena 
would be of some help e.g., to design future experiments. 
An example of such is the parameter degeneracy \cite{intrinsic,MNjhep01,octant}, 
the problem of multiple copy of the solutions of mixing parameters allowed by 
given sufficient but limited numbers of experimental data. 
See \cite{MU-Pdege} for a comprehensive overview of this phenomenon.
To facilitate understanding of qualitative features of the neutrino oscillation, 
it is crucially important to have analytic formula, albeit approximate, for the 
oscillation probability. For relatively short baseline experiments, 
such as low-energy superbeam \cite{MN-lowECP00,sato-01,richter-00}, 
the matter perturbation theory works \cite{AKS,MNprd98}. 
So far, most of the analyses for long baseline of $L \gsim 1000$ km were 
done by using the well known Cervera {\it et al.} formula \cite{golden}.\footnote{
However, it was shown that analysis of the parameter degeneracy with the 
matter-perturbative formula has been proved to give a transparent view of 
the phenomenon \cite{T2KK1,T2KK2,NSI-perturbation}, such as the decoupling 
between the degeneracies. 
}

A simple way of deriving the Cervera {\it et al.} formula is to expand the exact 
oscillation probability by small expansion parameters, 
$\epsilon \equiv \Delta m^2_{21} / \Delta m^2_{31}$ and $s_{13} \equiv \sin \theta_{13}$, 
both to second order. While the former is known to be small $\epsilon \simeq 0.03$, 
the latter can be larger by almost an order of magnitude; Currently, it is only 
bounded from above by the Chooz limit, $s_{13} \lsim 0.18$ 
\cite{CHOOZ,Palo-Verde,K2K13,MINOS13}. 
We note that there is an indication for nonzero $\theta_{13}$ comparable to 
the Chooz limit from global fit of the solar, reactor, atmospheric, and the 
accelerator experiments 
\cite{Fogli:2008jx,Schwetz:2008er,Maltoni:2008ka,GonzalezGarcia:2010er,KamLAND-13,SK-solar-III,schwetz-etal}.
Though the statistical significance of the indication is not high enough to be 
compelling, it certainly gives a good motivation for examining effects of such 
large values of $\theta_{13}$. 
Recently, a candidate event for $\nu_e$ appearance has been seen in the 
T2K experiment \cite{T2K-Mar16-2011} which further strengthens the motivation 
for taking the large-$\theta_{13}$ hypothesis seriously.

If $\theta_{13}$ is large, higher order terms of $s_{13}$ would be needed to 
achieve better agreement with the exact oscillation probability. 
In this paper, we compute such higher order corrections of $s_{13}$. 
To facilitate systematic computation we formulate a perturbative framework 
by assuming $s_{13}$ as large as $\simeq \sqrt{ \epsilon }$, 
which is comparable to the Chooz limit, 
and by taking the case of long enough baseline where the matter 
effect is comparable to the vacuum one. 
We call this framework as the ``$\sqrt{ \epsilon }$ perturbation theory'' as opposed 
to the $\epsilon$ perturbation theory for the Cervera {\it et al.} formula as named in 
\cite{NSI-perturbation}. We derive the second order formulas 
for the oscillation probabilities in all channels. 
By taking the ansatz, $s_{13} \simeq \sqrt{ \epsilon }$, our formulas 
will be valid for a wide range of $\theta_{13}$, 
$\epsilon \lsim s_{13} \lsim \sqrt{ \epsilon }$. 
The $\sqrt{ \epsilon }$ perturbation theory has been formulated earlier for relatively 
short baseline setting \cite{ustron09-mina}.

Some characteristic features of the computed formulas prompted us to think 
about general property of the oscillation probability, which resulted in the two 
general theorems described in Sec.~\ref{G-feature}. 
For example, we show that the $\delta$ dependent terms in the $\nu_e$-related 
oscillation probabilities exist only in odd terms of $s_{13}$, and conversely, 
all the terms odd in $s_{13}$ have $\delta$ dependence. 
Thanks to the theorems, quite (un)interestingly, we can prove on general ground 
that the correction terms to the Cervera {\it et al.} formula have to be second 
order in $\epsilon$, eliminating the possibility of having large corrections. 
Yet, we will observe that at super long baseline, $L=4000$ km, our correction 
terms nicely fill a sizable gap between the exact oscillation probability and the 
Cervera {\it et al.} formula, which exists in a limited range of energy for 
$\theta_{13}$ comparable to the Chooz limit.

The possible large value of $\theta_{13}$ could generate some interesting effects. 
For example, there arise terms of order $\epsilon^{5/2}$ consisting solely of 
$\delta$-dependent terms in the $\nu_{e}$ appearance oscillation probability,  
which is smaller only by a factor of $\simeq 5$ compared to the existing terms 
in the Cervera {\it et al.} formula.  
Another intriguing feature arises when the non-standard interactions (NSI) of 
neutrinos are included into the system. For an extensive list of the 
references for NSI including the original ones see, e.g., \cite{NOVE09-mina}.  
Some of the NSI dependent terms get enhanced by large $\theta_{13}$, and 
decoupling of some NSI elements from the $\nu_{e}$-related appearance 
probabilities no more hold.
On the other hand, the smallness of the correction terms to the Cervera {\it et al.} 
formula has a consequence that, quite naturally, their influence to parameter determination including the issue of parameter degeneracy is quite limited.

Following the discussion of general property of oscillation probability in 
Sec.~\ref{G-feature}, we formulate our $\sqrt{ \epsilon }$ perturbation theory 
(Sec.~\ref{formulation}), and derive the expressions of the oscillation probabilities 
valid to order $\epsilon^{2}$ in $\nu_{e} \rightarrow \nu_{\mu}$ and 
$\nu_{\mu} \rightarrow \nu_{\mu}$ channels (Sec.~\ref{oscillation-P}).
The characteristic features of the formulas are shown to be understood thanks to 
the theorems given in Sec.~\ref{G-feature}. The accuracy of our formula is 
checked against the exact results for a large value of $\theta_{13}$ (Sec.~\ref{accuracy}). 
Then, we calculate order $\epsilon^{5/2}$ terms of the oscillation probability in 
the $\nu_{e} \rightarrow \nu_\mu$ channel which could be relevant for 
measurement of $\delta$ in its extreme precision (Sec.~\ref{order-5/2}).
Implications of smallness of the large $\theta_{13}$ corrections are explored 
by treating the parameter degeneracy (Sec.~\ref{degeneracy}). 
The second-order formula of the oscillation probability in the $\nu_{e}$-related 
appearance channels is derived for systems with NSI  (Sec.~\ref{NSI}). 
Concluding remarks are given in Sec.~\ref{conclusion}. 
Appendices are devoted for a proof of a general theorem for suppression 
of CP phase effect (Appendix~\ref{proof-B}), the explicit expressions of the 
$S$ matrix elements in the $\theta_{23}$-rotated basis (Appendix~\ref{tilde-S}), 
and explicit forms of the oscillation probabilities in the remaining channels 
(Appendix~\ref{probability-app}).

\section{General Features of Neutrino Oscillations in Matter} 
\label{G-feature}

Before constructing perturbative framework of neutrino oscillation, let us discuss 
some generic features of the oscillation probability. 
They are interesting by themselves, and statements about $\cos \delta$ dependence 
do not appear to be explicitly spelled out in the literature to our knowledge. 
It will also help us to understand characteristic features of the perturbation theory 
of neutrino oscillation to be discussed in the rest of this paper. 
An example of this is that CP phase effect appear only in terms of half-integral order 
of the small expansion parameter $\epsilon$ in our $\sqrt{ \epsilon }$ 
perturbation theory.

We note that the general features of neutrino oscillation probability which exactly 
hold with \cite{KTY1,KTY2} or without \cite{KTY3} constant matter density approximation 
can be used to prove some useful properties of the oscillation probabilities. 
That is, we call the readers' attention to the following two theorems:

\begin{itemize}

\item 
Theorem A: In $\nu_e$-related oscillation probabilities the $\delta$-dependence 
exists only in odd terms in $s_{13}$. Conversely, all the odd terms in 
$s_{13}$ are accompanied with either $\cos \delta$ or $\sin \delta$.

\item 
Theorem B: $\delta$-dependent terms in the oscillation probability in matter, 
not only $\sin \delta$ but also $\cos \delta$ terms, must 
come with the two suppression factors, 
$\frac{ \Delta m^2_{21} }{ \Delta m^2_{31} }$ and 
the reduced (or ``not quite'') Jarlskog coefficient 
$J_{r} \equiv c_{12} s_{12} c_{23} s_{23} s_{13}$.

\end{itemize}

\noindent
In talking about $s_{13}$- and $\delta$-dependences we assume the standard 
form of the lepton flavor mixing matrix, the MNS matrix \cite{MNS} given in 
(\ref{MNSmatrix}), the convention which will be used throughout this paper.

To prove the theorem A we note that $s_{13}$ and $\delta$ enter into the Hamiltonian through the single variable $z \equiv s_{13} e^{i \delta}$. 
Therefore, the oscillation probability $P$ can be written as a power series expansion as 
$P = \sum_{n, m}^{\infty} f_{n m} z^{n} (z^{*})^m$, where $f_{n m} = f^{*}_{m n} $ 
for reality of $P$. 
On the other hand, the result obtained in \cite{KTY3} by extending discussions 
in \cite{solar-CP} says that there are only $\cos \delta$ and $\sin \delta$ terms in 
$\nu_e$ related oscillation probabilities, and no higher harmonics of $\delta$. 
It means that only the terms that satisfy $m =n \pm 1$ survive. 
It leaves the unique form of the oscillation probability 
$P = K( s^2_{13} ) s_{13} \cos \delta + M( s^2_{13} ) s_{13} \sin \delta$, 
where $K$ and $M$ are some functions. 
At the same time, this construction guarantees that all the odd terms in 
$s_{13}$ are accompanied by $\delta$.
This is nothing but the theorem A.

Now, let us discuss the theorem B. 
If the adiabaticity holds existence of the suppression factors by the $\Delta m^2$ 
ratio and the angle factors for $\sin \delta$ terms is well known. 
In fact, one can show that the latter is precisely the Jarlskog coefficient 
$J \equiv c_{12} s_{12} c_{23} s_{23} c^2_{13} s_{13}$ thanks to the Naumov identity \cite{naumov,harrison-scott} 
\begin{eqnarray} 
\bar{ \Delta m^2_{12} } \bar{\Delta m^2_{23} } \bar{\Delta m^2_{31} } 
\bar{J} \sin \bar{ \delta } &=& 
\Delta m^2_{12} \Delta m^2_{23} \Delta m^2_{31} J \sin \delta, 
\label{naumov}
\end{eqnarray}
where the quantities with over-bar in the left-hand side denote the ones in matter 
which correspond to the one in vacuum in the right-hand side of (\ref{naumov}). 
Notice that $\cos \bar{ \delta } $ does not contain $\sin \delta$ \cite{KTY2}, 
and the proportionality between $\sin \bar{ \delta }$ and  $\sin \delta$ is also 
guaranteed by the Toshev identity 
$\bar{ c } _{23} \bar{ s }_{23} \sin \bar{ \delta } = c_{23} s_{23} \sin \delta$ 
\cite{toshev-id}. 
The equation (\ref{naumov}) indicates that the $\sin \delta$ term vanishes 
in the limit of one of $\Delta m^2_{ij} \rightarrow 0$, or 
vanishing limit of one of the mixing angles.

Whether the same statement apply to $\cos \delta$ terms or not, or if the theorem 
B is valid at all for cases with generic matter density profiles, is not obvious. 
Apparently no general statement has been made in the literature. 
Vanishing of any $\delta$ dependence (including cosine) in the absence of 
$\Delta m^2_{21}$ can be proved similarly as the phase reduction theorem given 
in \cite{NSI-perturbation} as a special case of turning off the non-standard interactions. 
For constant matter density proportionality of $\cos \delta$ terms to the Jarlskog 
coefficient $J$ (reduced coefficient $J_{r}$) in the $\nu_{e}$-related channels 
(oscillation channels in the $\nu_{\mu} - \nu_{\tau}$ sector) is explicitly proved 
in \cite{KTY2} by deriving the exact forms of the oscillation probabilities.\footnote{
The difference of whether the suppression factor is provided by $J$ or $J_{r}$ 
does not make difference in our following discussions. Hence, we do not 
discuss this point further apart from making a brief comment in Appendix~\ref{proof-B}. 
}
%
Therefore, what is left is to show that the last statement holds under arbitrary matter 
density profile without recourse to the assumption of adiabaticity. The proof for this general case is described in Appendix~\ref{proof-B}.

\section{Formulating Large-$\theta_{13}$ Perturbation Theory}
\label{formulation}

In this section, we formulate a large-$\theta_{13}$ perturbation theory of neutrino 
oscillation. We use an ansatz 
\begin{eqnarray} 
s_{13} \simeq \sqrt{ \epsilon }, 
\hspace{8mm} 
\epsilon  \equiv \frac{ \Delta m^2_{21} }{ \Delta m^2_{31} }  \simeq 0.03
\label{ansatz}
\end{eqnarray}
to formulate our perturbative framework, hence the name 
``$\sqrt{ \epsilon }$ perturbation theory''. It implies 
$\sin^2 2\theta_{13} \simeq 0.12$, the value comparable with the Chooz limit. 
We work in anticipation of long baselines of several thousand kilometers, 
so that  $r_{A} \equiv a /  \Delta m^2_{31} \sim \mathcal{O} ( 1 )$, and 
$\Delta m^2_{31} L / E \sim \mathcal{O} ( 1 )$ to be not far from the first 
oscillation maximum. Hereafter, 
$a \equiv 2\sqrt{2} G_F N_e(x) E$ is a coefficient for measuring the matter effect 
on neutrinos propagating in medium of electron number density 
$N_e(x)$~\cite{wolfenstein}, 
where $G_F$ is the Fermi constant and $E$ is the neutrino energy. 
In perturbative calculation of the oscillation probability we take constant 
electron number density approximation.

If we take the ansatz $s_{13} \simeq \epsilon$ (which corresponds to 
$\sin^2 2\theta_{13} \simeq 4 \epsilon^2 \simeq 4 \times 10^{-3}$) instead of 
(\ref{ansatz}),  
we obtain the widely used Cervera {\it et al.} formula \cite{golden} by keeping 
terms to order $\epsilon^2$. 
Of course, there may exist the other small (or large) parameters, such as 
$\Delta m^2_{31} L / E$ and $a /  \Delta m^2_{31}$ at short baselines, 
depending upon the experimental settings. 
The similar $\sqrt{ \epsilon }$ perturbation theory for shorter baselines is 
discussed by assuming $r_{A} \sim \sqrt{ \epsilon }$ \cite{ustron09-mina}.

The remaining potentially small parameters would be $\pi/4 - \theta_{23}$, but 
we do not take it as an expansion parameter for two reasons:\footnote{
Nonetheless, one can of course further expand the oscillation probability 
in terms of $\pi/4 - \theta_{23}$ assuming it small, as is done e.g., 
in \cite{pakvasa07,altarelli-meloni,NSI-Madrid}. 
}
(1) A rather large range is currently allowed for $\theta_{23}$, 
and moreover the situation will not be changed even with the next 
generation experiments \cite{MSS04}.
(2) As will become evident in formulating our perturbative framework $\theta_{23}$ 
is an ``external parameter'' which is irrelevant in doing perturbative computation.

We follow \cite{NSI-perturbation} to formulate the 
perturbative treatment of neutrino oscillation. 
The $S$ matrix describes possible flavor changes after traversing 
a distance $L$, 
\begin{eqnarray} 
\nu_{\alpha} (L) = S_{\alpha \beta} \nu_{\beta} (0), 
\label{def-S}
\end{eqnarray}
and the oscillation probability is given by 
\begin{eqnarray} 
P(\nu_{\beta} \rightarrow \nu_{\alpha}; L )= 
\vert S_{\alpha \beta} \vert^2.  
\label{def-P}
\end{eqnarray}
When the neutrino evolution is governed by the Schr\"odinger equation, 
$ i \frac{d}{dx} \nu = H \nu  $, 
$S$ matrix is given as 
\begin{eqnarray} 
S = T \text{exp} \left[ -i \int^{L}_{0} dx H(x) \right] 
\label{evolution}
\end{eqnarray}
where $T$ symbol indicates the ``time ordering''  
(in fact ``space ordering'' here). 
The right-hand side of (\ref{evolution}) may be written as 
$e^{-i H L}$ for the case of constant matter density. 
In the standard three-flavor neutrinos, Hamiltonian is given 
(with $a = 2\sqrt{2} G_F N_e E$) by 
\begin{eqnarray}
H= 
\frac{ 1 }{ 2E } \left\{ 
U \left[
\begin{array}{ccc}
0 & 0 & 0 \\
0 & \Delta m^2_{21}& 0 \\
0 & 0 & \Delta m^2_{31} 
\end{array}
\right] U^{\dagger}
+
\left[
\begin{array}{ccc}
a(x) & 0 & 0 \\
0 & 0 & 0 \\
0 & 0 & 0
\end{array}
\right] 
\right\}, 
\label{hamiltonian}
\end{eqnarray}
where $\Delta m^2_{ji} \equiv m^2_{j} - m^2_{i}$. 
In (\ref{hamiltonian}) $U$ is the MNS matrix and can be written in the standard notation as 
\begin{eqnarray}
U = U_{23} U_{13} U_{12} = 
\left[
\begin{array}{ccc}
1 & 0 &  0  \\
0 & c_{23} & s_{23} \\
0 & - s_{23} & c_{23} \\
\end{array}
\right] 
\left[
\begin{array}{ccc}
c_{13}  & 0 &  s_{13} e^{- i \delta}   \\
0 & 1 & 0 \\
- s_{13} e^{ i \delta}  & 0 & c_{13}  \\
\end{array}
\right] 
\left[
\begin{array}{ccc}
c_{12} & s_{12}  &  0  \\
- s_{12} & c_{12} & 0 \\
0 & 0 & 1 \\
\end{array}
\right] 
\label{MNSmatrix}
\end{eqnarray}
with the notation $s_{ij} \equiv \sin \theta_{ij}$ etc. and 
$\delta$ being the lepton KM phase.

To formulate perturbative treatment it is convenient to work with 
the tilde basis defined as 
$\tilde{\nu}_{\alpha} = (U_{23}^{\dagger})_{\alpha \beta} \nu_{\beta}$, 
in which the Hamiltonian is related to 
the flavor basis one as \cite{munich04} 
\begin{eqnarray} 
\tilde{H} = U_{23}^{\dagger} H U_{23}
\label{tilde-hamiltonian}
\end{eqnarray}
where $U_{23}$ is defined in (\ref{MNSmatrix}). 
The $S$ matrix in the flavor basis is related to the $S$ matrix in the tilde basis as 
\begin{eqnarray} 
S(L) = U_{23} \tilde{S} (L) U_{23}^{\dagger}
\label{Smatrix}
\end{eqnarray}
where 
$ \tilde{S} (L) = T \text{exp} \left[ -i \int^{L}_{0} dx \tilde{H} (x)  \right] $. 
The unperturbed part of the tilde-basis Hamiltonian is given by 
\begin{eqnarray} 
\tilde{H}_{0} &=& \Delta 
\left[
\begin{array}{ccc}
r_{A} & 0 & 0 \\
0 & 0 & 0 \\
0 & 0 & 1
\end{array}
\right], 
\label{H0}
\end{eqnarray}
where $\Delta \equiv \frac{ \Delta m^2_{31} }{ 2E }$ and 
$r_{A} \equiv \frac{ a }{ \Delta m^2_{31} }$. 
While the perturbed part is written with another simplified notation 
$r_{ \Delta } \equiv \frac{ \Delta m^2_{21} }{ \Delta m^2_{31} }$ as 
\begin{eqnarray} 
\tilde{H}_{1} &=& 
\Delta 
\left[
\begin{array}{ccc}
0 & 0 & s_{13} e^{ -i \delta} \\
0 & 0 & 0 \\
s_{13} e^{ i \delta} & 0 & 0 
\end{array}
\right] 
+
\Delta  \left[
\begin{array}{ccc}
r_{\Delta} s^2_{12} + s^2_{13} & r_{\Delta}  c_{12} s_{12}  & 0 \\
r_{\Delta}  c_{12} s_{12}  & r_{\Delta}  c^2_{12}  & 0 \\
0 & 0 & - s^2_{13}  
\end{array}
\right] 
\nonumber \\
&-&  
\Delta 
\left[
\begin{array}{ccc}
0  & 0 & \left( r_{\Delta}  s^2_{12} + \frac{1}{2} s^2_{13} \right) s_{13} e^{ -i \delta}  \\
0 & 0 & r_{\Delta} c_{12} s_{12} s_{13} e^{ -i \delta}   \\
\left( r_{\Delta}  s^2_{12} + \frac{1}{2} s^2_{13} \right) s_{13} e^{ i \delta}  & r_{\Delta} c_{12} s_{12} s_{13} e^{ i \delta}  & 0  
\end{array}
\right] 
\nonumber \\ 
&-&
\Delta r_{\Delta}  \left[
\begin{array}{ccc}
 s^2_{12} s^2_{13} & \frac{1}{2} c_{12} s_{12} s^2_{13} & 0 \\
 \frac{1}{2} c_{12} s_{12} s^2_{13} & 0  & 0 \\
0 & 0 & - s^2_{12} s^2_{13}
\end{array}
\right]. 
\label{H1}
\end{eqnarray}
The first, second, third, and the fourth terms in (\ref{H1}) are of order 
$\epsilon^{\frac{1}{2}}$, 
$\epsilon^{1}$, 
$\epsilon^{\frac{3}{2}}$, and 
$\epsilon^{2}$, respectively.

To calculate $ \tilde{S} (L)$ perturbatively we define $\Omega(x)$ as 
$\Omega(x) = e^{i \tilde{H}_{0} x} \tilde{S} (x)$, which obeys the evolution equation 
\begin{eqnarray} 
i \frac{d}{dx} \Omega(x) = H_{1} \Omega(x) 
\label{omega-evolution}
\end{eqnarray}
where 
\begin{eqnarray} 
H_{1} \equiv e^{i \tilde{H}_{0} x} \tilde{H}_{1} e^{-i \tilde{H}_{0} x} 
\label{def-H1}
\end{eqnarray}
Then, $\Omega(x)$ can be computed perturbatively as 
\begin{eqnarray} 
\Omega(x) &=& 1 + 
(-i) \int^{x}_{0} dx' H_{1} (x') + 
(-i)^2 \int^{x}_{0} dx' H_{1} (x') \int^{x'}_{0} dx'' H_{1} (x'') 
\nonumber \\
&+& (-i)^2 \int^{x}_{0} dx' H_{1} (x') \int^{x'}_{0} dx'' H_{1} (x'') \int^{x''}_{0} dx''' H_{1} (x''') + 
\mathcal{O} ( \epsilon^4 ). 
\label{Omega-exp}
\end{eqnarray}
where the ``space-ordered'' form in (\ref{Omega-exp}) is essential 
because of the non-commutativity between $H_{1}$ of different locations. 
Having obtained $\Omega(x)$ $\tilde{ S }$ matrix can be written as 
\begin{eqnarray} 
\tilde{ S }(x) = e^{- i \tilde{H}_{0} x} \Omega(x). 
\label{Smatrix-tilde2}
\end{eqnarray}
The results of $ \tilde{S} (x)$ matrix elements to second-order in $\epsilon$ 
are given in Appendix~\ref{tilde-S}.
Then, the $S$ matrix can be computed by making a rotation in (2-3) space 
$S = U_{23} \tilde{ S } U_{23}^{\dagger}$ as in (\ref{Smatrix}). (See (\ref{SbyS-tilde}).)
Finally, the oscillation probability can readily be obtained by using (\ref{def-P}).
For example, the one in the $\nu_{e} \rightarrow \nu_\mu$ channel 
can be given by using the $\tilde{S}$ matrix elements as 
%
\begin{eqnarray}
P (\nu_e \rightarrow \nu_\mu) &\equiv& 
\vert S_{\mu e} \vert^2 = 
s^2_{23} \vert \tilde{S}^{(1/2)}_{e \tau} (-\delta) \vert^2 + 
2 c_{23} s_{23} \text{Re} \left[
\tilde{S}^{(1/2)}_{e \tau} (-\delta)^{*} \tilde{S}^{(1)}_{e \mu} (-\delta) 
\right] 
\nonumber \\ 
&+& 
c^2_{23} \vert \tilde{S}^{(1)}_{e \mu} (-\delta) \vert^2 +  
2 s^2_{23} \text{Re} \left[
\tilde{S}^{(1/2)}_{e \tau} (-\delta)^{*} \tilde{S}^{(3/2)}_{e \tau} (-\delta)
\right]. 
\label{Pemu-by-tildeS}
\end{eqnarray}

\section{Perturbative Expression of the Oscillation Probability} 
\label{oscillation-P}

In this section, we present perturbative expressions of the oscillation probabilities 
in the $\nu_{e}$ related and the $\nu_{\mu}-\nu_{\tau}$ sectors to order 
$\epsilon^2$ in our $\sqrt{ \epsilon }$ perturbation theory. 
All the formulas in this section are given in the form as 
\begin{eqnarray}
P (\nu_\alpha \rightarrow \nu_\beta) &=& 
P_{\alpha \beta}^{(0)} + P_{\alpha \beta}^{(1)} + 
P_{\alpha \beta}^{(3/2)} + P_{\alpha \beta}^{(2)} 
\label{Palpha-beta-def}
\end{eqnarray}
in which we use $L$ to denote the baseline distance.
Some order $\epsilon^{5/2}$ terms will be discussed in Sec.~\ref{order-5/2}. 
The antineutrino probability can be obtained from the neutrino probability by 
the replacement as 
$P (\bar{\nu}_\alpha \rightarrow \bar{\nu}_\beta; \delta, a) = 
P (\nu_\alpha \rightarrow \nu_\beta; - \delta, -a)$. 
Similarly, the T-conjugate one is given by 
$P (\nu_\beta \rightarrow \nu_\alpha; \delta, a) = 
P (\nu_\alpha \rightarrow \nu_\beta; - \delta, a)$. 
Remember that the following abbreviated notations are used: 
$\Delta \equiv \frac{ \Delta m^2_{31} }{ 2E }$, 
$r_{\Delta} \equiv \frac{  \Delta m^2_{21} }{ \Delta m^2_{31} }$, 
 and 
$r_{A} \equiv \frac{ a }{ \Delta m^2_{31} }$.

\subsection{Oscillation Probabilities in $\nu_{e}$ related sector} 
\label{e-alpha}

There is no zeroth oder term in the appearance channels. 
With use of the reduced Jarlskog coefficient 
$J_{r} \equiv c_{12} s_{12} c_{23}  s_{23} s_{13}$, the order 
$\epsilon^{1}$, $\epsilon^{3/2}$, and $\epsilon^{2}$ terms in the 
oscillation probability in the $\nu_{e} \rightarrow \nu_\mu$ channel are given by 
%
\begin{eqnarray}
P_{e \mu}^{(1)} &=& 
4 s^2_{23} s^2_{13} \frac{ 1 }{ (1 - r_{A})^2 } 
\sin^2 \frac{ (1 - r_{A}) \Delta L }{ 2 }, 
\label{Pemu-1}
\end{eqnarray}
%
\begin{eqnarray}
P_{e \mu}^{(3/2)} &=& 
8 J_{r} \frac{ r_{\Delta} }{ r_{A} (1 - r_{A}) }
\cos \left( \delta - \frac{ \Delta L }{ 2 } \right)
\sin \frac{ r_{A} \Delta L }{ 2 } 
\sin \frac{ (1 - r_{A}) \Delta L }{ 2 },   
\label{Pemu-3/2}
\end{eqnarray}
%
\begin{eqnarray}
P_{e \mu}^{(2)} &=& 
4 c^2_{23} c^2_{12} s^2_{12} 
\left( \frac{ r_{\Delta} }{ r_{A} } \right)^2 
\sin^2 \frac{ r_{A} \Delta L }{ 2 } 
\nonumber \\
&-& 
4 s^2_{23} \left[ 
s^4_{13} \frac{ (1 + r_{A})^2 }{ (1 - r_{A})^4 } 
- 2 s^2_{12} s^2_{13} \frac{ r_{\Delta} r_{A} }{ (1 - r_{A})^3 } 
\right] 
\sin^2 \frac{ (1 - r_{A}) \Delta L }{ 2 } 
\nonumber \\
&+& 
2 s^2_{23} \left[ 
2 s^4_{13} \frac{ r_{A} }{ (1 - r_{A})^3 } 
- s^2_{12} s^2_{13} \frac{ r_{\Delta} }{ (1 - r_{A})^2 } 
\right] 
(\Delta L ) \sin (1 - r_{A}) \Delta L.
\label{Pemu-2}
\end{eqnarray}

As is obvious from (\ref{SbyS-tilde}), the similar expressions $P_{e \tau}^{(i)}$ 
for the $\nu_{e} \rightarrow \nu_\tau$ channel can be obtained from 
$P_{e \mu}^{(i)}$ in (\ref{Pemu-1})-(\ref{Pemu-2}) by the transformation 
$c_{23} \rightarrow - s_{23}$ and $s_{23} \rightarrow c_{23}$. 
The explicit expressions are given in Appendix~\ref{probability-app}.
Given $P_{e \mu}^{(i)}$ and $P_{e \tau}^{(i)}$, $P_{ee}^{(i)}$ can be obtained 
by using the perturbative unitarity relation\footnote{
In fact, we computed all the $P_{e \alpha}^{(i)}$ ($\alpha = e, \mu, \tau$) by the same procedure and explicitly verified the perturbative unitarity relation.
}
%
\begin{eqnarray}
P_{ee}^{(i)} = \delta_{i, 0} -  P_{e \mu}^{(i)} -  P_{e \tau}^{(i)} 
\hspace{10mm} 
(i = 0, 1, 3/2, 2). 
\label{P-unitarity}
\end{eqnarray}
Therefore, we do not present their explicit forms. 
Notice that $P_{e \mu}^{(3/2)} + P_{e \tau}^{(3/2)} = 0$ as it should because there must be no $\delta$ dependent terms in $P_{ee}$ \cite{kuo,solar-CP}.

\subsection{Oscillation Probabilities in $\nu_{\mu} - \nu_{\tau}$ sector} 
\label{mu-tau}

The order-by-order perturbative formulas of the oscillation probabilities in the 
$\nu_\mu \rightarrow \nu_\mu$ channel to order $\epsilon^{2}$ can be computed 
via a similar manner. The results are given by 
%
\begin{eqnarray}
P_{\mu \mu}^{(0)} &=& 
1 - 4 c^2_{23}  s^2_{23} \sin^2 \left( \frac{ \Delta L }{ 2 } \right), 
\label{Pmumu-0}
\end{eqnarray}
\begin{eqnarray}
P_{\mu \mu}^{(1)} &=& 
- 4 s^4_{23} s^2_{13} \frac{ 1 }{ (1 - r_{A})^2 } 
\sin^2 \frac{ (1 - r_{A}) \Delta L }{ 2 } 
\nonumber \\
&-& 
2 c^2_{23}  s^2_{23} 
\left[
s^2_{13}  \frac{ r_{A} }{ 1 - r_{A} } - c^2_{12} r_{\Delta} 
\right] 
( \Delta L ) \sin \Delta L 
\nonumber \\
&+& 
4 c^2_{23}  s^2_{23} s^2_{13} \frac{ 1 }{ (1 - r_{A})^2 } 
\sin \frac{ (1 + r_{A}) \Delta L }{ 2 } \sin \frac{ (1 - r_{A}) \Delta L }{ 2 }, 
\label{Pmumu-1}
\end{eqnarray}
\begin{eqnarray}
P_{\mu \mu}^{(3/2)} &=& 
- 8 J_{r} \cos \delta \left( c^2_{23} - s^2_{23} \right) 
\frac{ r_{\Delta} r_{A} }{ 1 - r_{A}  }
\sin^2 \left( \frac{ \Delta L }{ 2 } \right) 
\nonumber \\
&+& 
8 J_{r} \cos \delta 
\frac{ r_{\Delta} }{ r_{A} (1 - r_{A} ) }
\sin^2 \left( \frac{ r_{A} \Delta L }{ 2 } \right) 
\nonumber \\
&-& 
16 J_{r} \cos \delta~s^2_{23} 
\frac{ r_{\Delta} }{ r_{A} (1 - r_{A} ) }
\sin \frac{ \Delta L }{ 2 } 
\sin \frac{ r_{A} \Delta L }{ 2 } 
\cos \frac{ (1 - r_{A}) \Delta L }{ 2 }, 
\label{Pmumu-3/2}
\end{eqnarray}
\begin{eqnarray}
&& P_{\mu \mu}^{(2)} = 
- c^2_{23}  s^2_{23} 
\left( s^2_{13} \frac{ r_{A} }{ 1 - r_{A} } - c^2_{12} r_{\Delta}  \right)^2 
( \Delta L )^2 \cos \Delta L 
\nonumber \\
&+& 
2 c^2_{23}  s^2_{23} 
\left[
s^4_{13}  \frac{ r_{A} (1 + r_{A}) }{ (1 - r_{A})^3 } 
- (c^2_{12} + s^2_{12} r_{A}^2) s^2_{13} \frac{ r_{\Delta} }{ (1 - r_{A})^2 } 
- c^2_{12} s^2_{12}  \frac{ r_{\Delta}^2 }{ r_{A} }
\right] 
( \Delta L ) \sin \Delta x
\nonumber \\
&+& 
2 c^2_{23}  s^2_{23} 
\left[ 
s^4_{13}  \frac{ r_{A} }{ (1 - r_{A})^3 } 
+ (c^2_{12} - s^2_{12}) s^2_{13} \frac{ r_{\Delta} }{ (1 - r_{A})^2 } 
\right] 
( \Delta L ) \sin r_{A} \Delta L 
\nonumber \\
&-& 
2 s^4_{23} 
\left[ 
2 s^4_{13}  \frac{ r_{A} }{ (1 - r_{A})^3 } 
- s^2_{12} s^2_{13} \frac{ r_{\Delta} }{ (1 - r_{A})^2 } 
\right] 
( \Delta L ) \sin (1 - r_{A}) \Delta L 
\nonumber \\
&-& 
4 c^2_{23}  s^2_{23} 
\left[
s^4_{13} \frac{ r_{A} ( 2 + r_{A} ) }{ (1 - r_{A})^4 }
- 2 s^2_{12}  s^2_{13} \frac{ r_{\Delta} r_{A} }{ (1 - r_{A})^3 } 
- c^2_{12} s^2_{12} \left( \frac{ r_{\Delta} }{ r_{A} } \right)^2 
\right] 
\sin^2 \frac{ \Delta L }{ 2 } 
\nonumber \\
&+& 
4 
\left[
c^2_{23}  s^2_{23} s^4_{13} \frac{ r_{A} ( 2 + r_{A} ) }{ (1 - r_{A})^4 }
- 2 c^2_{23}  s^2_{23} s^2_{12}  s^2_{13} \frac{ r_{\Delta} r_{A} }{ (1 - r_{A})^3 } 
- c^4_{23} c^2_{12} s^2_{12} \left( \frac{ r_{\Delta} }{ r_{A} } \right)^2 
\right] 
\sin^2 \frac{ r_{A} \Delta L }{ 2 } 
\nonumber \\
&+& 
4 
\left[
s^4_{23} s^4_{13} \frac{ ( 1 + r_{A} )^2 }{ (1 - r_{A})^4 }
- 2 s^4_{23} s^2_{12}  s^2_{13} \frac{ r_{\Delta} r_{A} }{ (1 - r_{A})^3 } 
- c^2_{23}  s^2_{23} c^2_{12} s^2_{12} \left( \frac{ r_{\Delta} }{ r_{A} } \right)^2 
\right] 
\sin^2 \frac{ (1- r_{A}) \Delta L }{ 2 }. 
\nonumber \\
\label{Pmumu-2}
\end{eqnarray}
%

The similar order $\epsilon^{1}$, $\epsilon^{3/2}$, and $\epsilon^{2}$ terms 
$P_{\mu \tau}^{(i)}$ in the oscillation probability in the 
$\nu_\mu \rightarrow \nu_\tau$ channel are given in Appendix~\ref{probability-app}.
For $\nu_\mu \rightarrow \nu_e$ channel, 
$P_{\mu e}^{(i)} (\delta)$ ($i=0,1, 3/2, 2$) can be obtained from its T-conjugate 
as $P_{e \mu}^{(i)} (-\delta)$. 
It is then straightforward to verify that the similar perturbative unitarity relation, with 
the indices $e$ being replaced by $\mu$ in (\ref{P-unitarity}), holds, the task we have 
explicitly executed.

\subsection{Understanding some characteristic features of the perturbative formulas}
\label{understanding}

We first note that the $\delta$ dependence appears only in terms with half-integral 
order in $\epsilon$, and all the terms of half integral order in $\epsilon$ contains 
$\delta$ dependence. It can be readily understood by the theorem A 
in Sec.~\ref{G-feature} because odd terms of $s_{13}$ have to be half-integral 
order of $\epsilon$ in our $\sqrt{ \epsilon }$ perturbation theory.

Despite that the Cervera {\it et al.} formula is the second order formula for small 
$s_{13} \sim \epsilon$, practically it is often used even for relatively large $\theta_{13}$, 
for example, in the analysis of parameter degeneracy \cite{MNP2,MU-Pdege}. 
Therefore, it is a legitimate question to ask how large the corrections terms to the 
formula can be for large $\theta_{13}$. 
We therefore classify each term in the oscillation probabilities into the two 
categories, the one which exist in the Cervera {\it et al.} formula (denoted for brevity 
as the Cervera terms), and the ones which do not (non-Cervera terms). 
The non-Cervera terms in each channel are the terms with factors of either 
$s^4_{13}$ or $r_{\Delta} s^2_{13}$. 
In the $\nu_{e} \rightarrow \nu_{\mu}$ ($\nu_{e} \rightarrow \nu_{\tau}$) channel 
they are the last two lines in (\ref{Pemu-2}) ((\ref{Petau-2})), which come from the 
last term in (\ref{Pemu-by-tildeS}).

One notices that the non-Cervera terms arise only from the order $\epsilon^2$ terms. 
To understand this feature let us first note that there can be no non-Cervera terms 
of order $\epsilon^{1}$ in our $\sqrt{ \epsilon }$ perturbation theory. 
While the $\epsilon^{1}$ suppression is provided by the factor of either $r_{ \Delta }$ 
or $s^2_{13}$, they can be among the $\epsilon^{2}$ terms in the $\epsilon$ 
perturbation theory, which is included in the Cervera terms. 
Therefore, the largest possible non-Cervera terms may be contained in the 
order $\epsilon^{3/2}$ terms. 
However, they {\em do not} exist for the following reason: 
The theorem A proved in Sec.~\ref{G-feature} states that all the odd terms in 
$s_{13}$ must be accompanied either by $\cos \delta$ or $\sin \delta$. 
Then, the theorem B dictates that all the $\delta$ dependent terms must 
have extra suppression factor 
$r_{ \Delta } \equiv \Delta m^2_{21} / \Delta m^2_{31} \sim \epsilon$. 
Therefore, $s^3_{13}$ terms in the non-Cervera part of the oscillation probability 
are actually of order $\epsilon^{5/2}$.
As a consequence, the non-Cervera terms exist in our second-order formulas 
only with the suppression factors of $s^4_{13}$ or $r_{ \Delta } s^2_{13}$. 
Hence, they are of order $\epsilon^{2}$, excluding the possibility of yielding 
large corrections of order $\epsilon^{3/2}$ or lower from the non-Cervera terms.

\section{Accuracy of the approximate formula for large $\theta_{13}$}
\label{accuracy}

\begin{figure}[htbp]
\begin{center}
\includegraphics[width=0.84\textwidth]{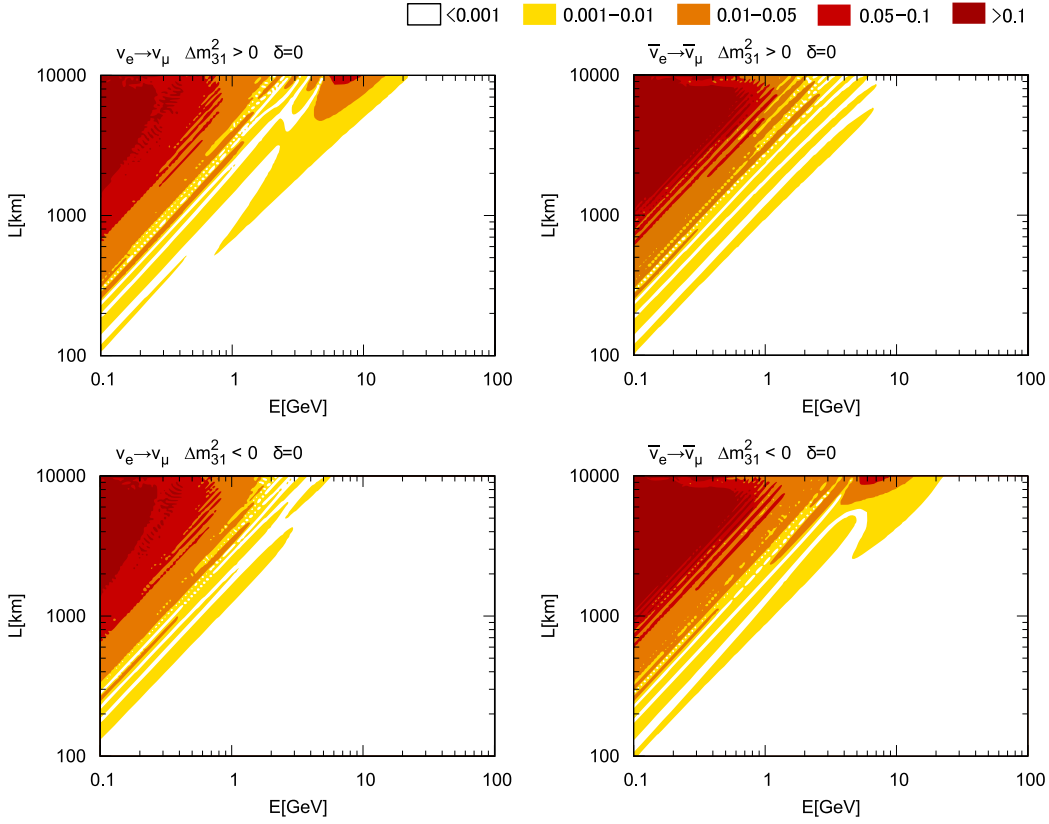}
\vspace{-2mm}
\end{center}
\caption{
The absolute difference between the numerically evaluated exact 
oscillation probability and our approximate formula to order $\epsilon^2$ in the 
$\nu_{e} \rightarrow \nu_{\mu}$ channel, 
$\vert P_{e \mu}^{ \text{exact} } -  P_{e \mu}^{ \text{2nd} } \vert$, 
is presented by using the color graduation plot in $E-L$ space. 
The top (bottom) two panels are for the normal (inverted) mass hierarchy, 
the left neutrino and the right anti-neutrino channels.
The correspondence between colors and the probability difference 
is given at the top of the figure. 
The matter density is taken as 3.0 g/cm$^3$ for all baselines. 
$\theta_{13}$ is taken as $\sin \theta_{13} = 0.18$ and $\delta =0$. 
The remaining mixing parameters are chosen as 
$\Delta m^2_{31} = 2.4 \times 10^{-3}$ eV$^{2}$, 
$\Delta m^2_{21} = 7.7 \times 10^{-5}$ eV$^{2}$, 
$\sin^2 \theta_{23} = 0.5$, and 
$\tan^2 \theta_{12} = 0.44$. 
}
\label{difference-EL}
\end{figure}

\begin{figure}[htbp]
\begin{center}
\includegraphics[width=0.84\textwidth]{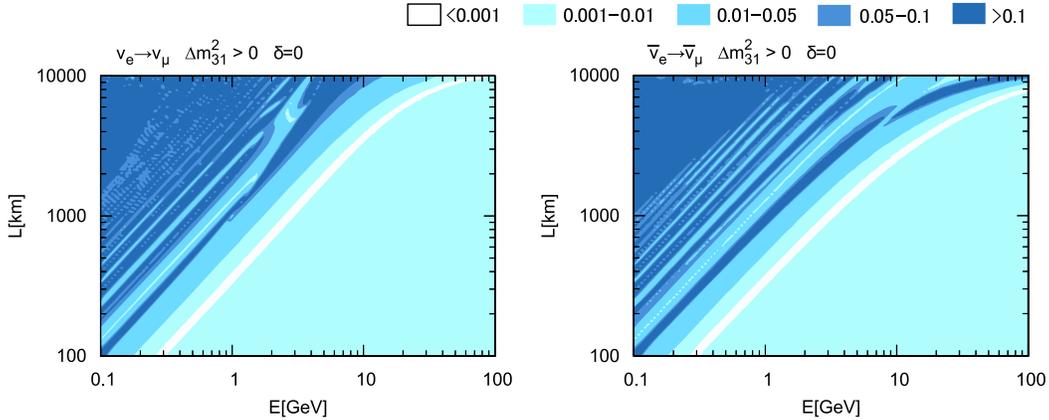}
\vspace{-2mm}
\end{center}
\caption{The relative difference 
$\vert P_{e \mu}^{ \text{exact} } -  P_{e \mu}^{ \text{2nd} } \vert / P_{e \mu}^{ \text{exact} }$  
is presented with the same format and by using the same values of the 
parameters as in Fig.~\ref{difference-EL}. 
}
\label{ratio-EL}
\end{figure}

\begin{figure}[htbp]
\begin{center}
\includegraphics[width=0.42\textwidth]{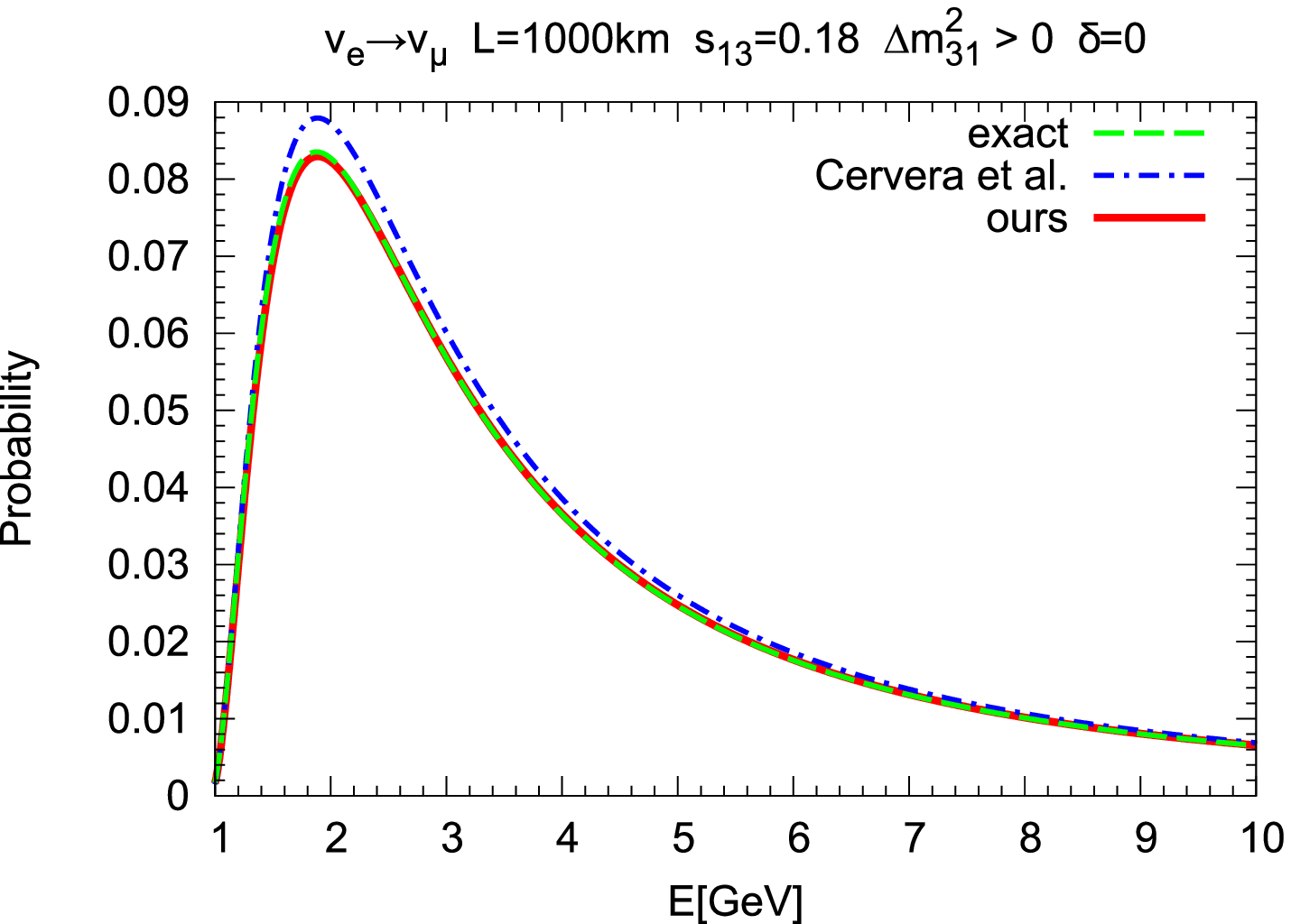}
\includegraphics[width=0.42\textwidth]{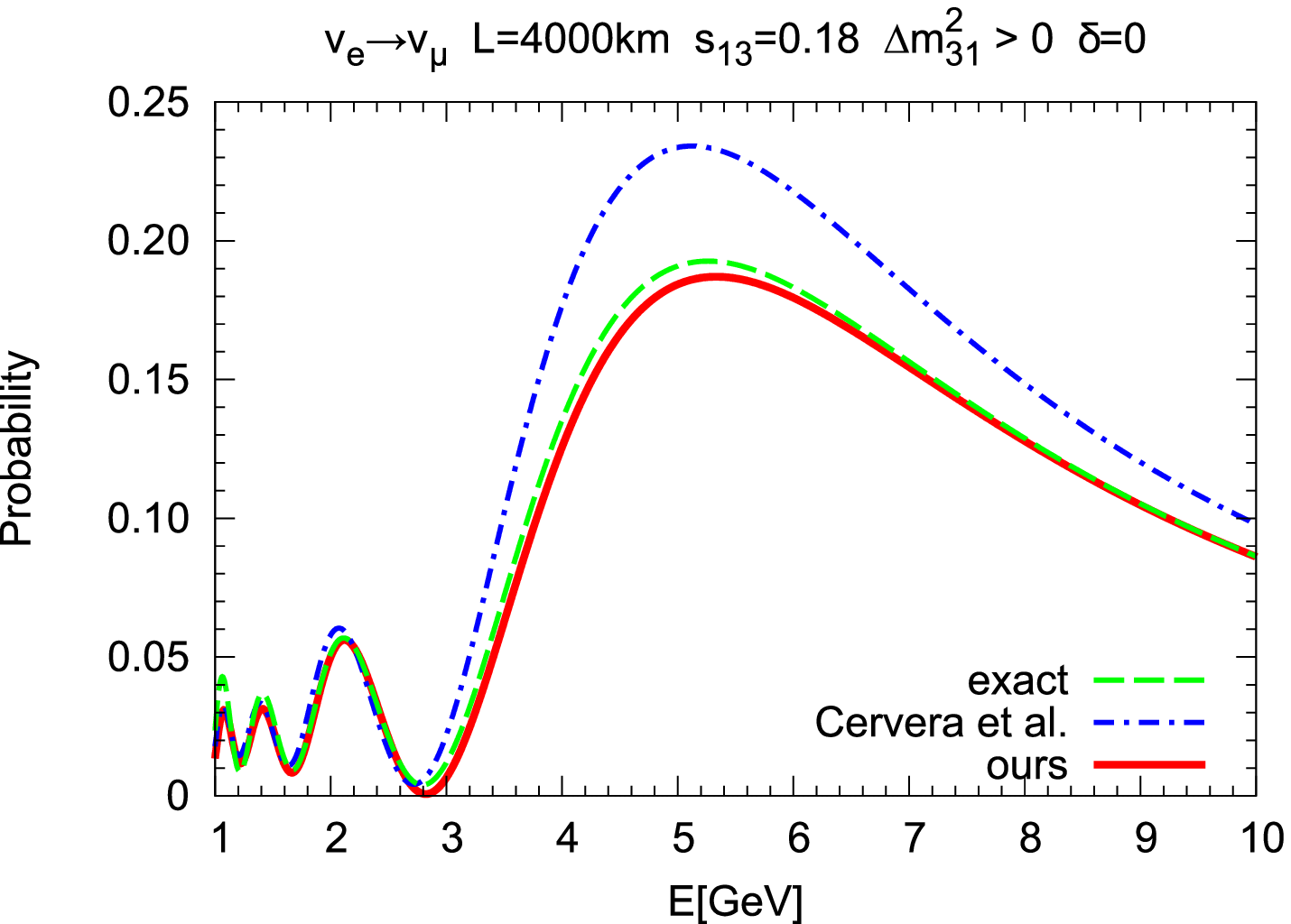}
\end{center}
\caption{
Comparison between the exact oscillation probability $P (\nu_{e} \rightarrow \nu_{\mu})$ 
computed numerically as a function of energy (green dashed line), 
the one calculated by the Cervera {\it et al.} formula (blue dash-dotted line), 
and with our formula with large $\theta_{13}$ corrections (red solid line). 
The left and the right panels are for baselines $L=1000$ km and $L=4000$ km 
for which the matter density is taken as 2.8 g/cm$^3$ and 3.6 g/cm$^3$, respectively.
$\theta_{13}$ is taken as $\sin \theta_{13} = 0.18$ and $\delta =0$. 
The values of the remaining mixing parameters are the same as given in the 
caption of Fig.~\ref{difference-EL}. 
}
\label{Pemu_1000-4000}
\end{figure}

In this section, we examine numerical accuracy of our perturbative formulas of 
the oscillation probabilities computed to order $\epsilon^{2}$  for $\theta_{13}$ 
of the order of the Chooz limit. 
We focus on the $\nu_{e} \rightarrow \nu_{\mu}$ channel whereas we also did the 
similar analysis in the  $\nu_{\mu} \rightarrow \nu_{\mu}$ channel \cite{asano-thesis}. 
To display accuracy of the approximate formula, two typical ways of plot are available, namely, the absolute difference 
$\vert P_{e \mu}^{ \text{exact} } -  P_{e \mu}^{ \text{2nd} } \vert$, 
and the relative difference 
$\vert P_{e \mu}^{ \text{exact} } -  P_{e \mu}^{ \text{2nd} } \vert / P_{e \mu}^{ \text{exact} }$, where $P_{e \mu}^{ \text{exact} }$ and $P_{e \mu}^{ \text{2nd} }$ denote, respectively, 
numerically evaluated exact oscillation probability and our approximate formula 
to order $\epsilon^2$.  

In Figs.~\ref{difference-EL} and \ref{ratio-EL}, the absolute and the relative 
differences, respectively, are presented by using the color graduation plots 
in $E-L$ space. 
We note that the comparison between the Cervera {\it et al.} formula and the exact 
result was performed in \cite{munich04}. 
We use the similar format to make easier the comparison between our and their results. 
A comparison between the relevant panels in Figs.~3 and 4 of \cite{munich04} and 
Figs.~\ref{difference-EL} and \ref{ratio-EL} indicates that our large-$\theta_{13}$ correction terms 
improves the accuracy of the approximate formula in a wide region in $E-L$ space. 
It is even more so considering that our assumed value of $\sin \theta_{13}$ is 
about a factor of two larger than their largest value. 
The features of the plots with other values of $\delta$ are quite similar to those 
presented in Figs.~\ref{difference-EL} and \ref{ratio-EL}, and hence we do not 
present them. For the same reason, only the case of normal mass hierarchy 
is shown in Fig.~\ref{ratio-EL}.

To see more clearly how good (or bad) are the approximations by our and the 
Cervera {\it et al.} formula, we present in Fig.~\ref{Pemu_1000-4000} the oscillation 
probability as a function of neutrino energy 
calculated numerically (denoted as ``exact'', green dashed line), 
computed by using the Cervera {\it et al.} formula (blue dash-dotted line), 
computed with our perturbative formula (red solid line). 
The left and the right panels in Fig.~\ref{Pemu_1000-4000} are for baselines 
$L=1000$ km and $L=4000$ km, for which the matter density is taken as 
2.8 g/cm$^3$ and 3.6 g/cm$^3$, respectively. 
The mass hierarchy is taken as the normal one, $\Delta m^2_{31} > 0$. 
$\theta_{13}$ is taken as $\sin \theta_{13} = 0.18$ which is close to the Chooz limit, 
and $\delta =0$. The values of the remaining mixing parameters used are given 
in the caption of Fig.~\ref{difference-EL}. 

As seen in Fig.~\ref{Pemu_1000-4000}, the difference between the exact and the Cervera {\it et al.} formula is very visible (modest) at $L=4000$ km ($L=1000$ km). 
It is notable that the higher order corrections of $s_{13}$ incorporated in our formula nicely fill the gap between them. 
To understand the nature of the gap we have examined the other cases of mass hierarchy and (anti-)neutrino channel. 
The features of the three curves presented in Fig.~\ref{Pemu_1000-4000} are 
very similar to the ones of antineutrino channel with the inverted hierarchy, both 
$r_{A} \equiv \frac{a}{\Delta m^2_{31} }> 0$.
The features in the other two cases, antineutrino channel with the normal hierarchy 
and neutrino channel with the inverted hierarchy, both $r_{A} < 0$, are similar 
to each other. They are not presented because the differences among the three 
curves are much smaller for both $L=1000$ km and 4000 km. 
We have checked that the features mentioned above are very similar for 
other values of $\delta$.

As was proved in Sec.~\ref{understanding} the higher order corrections 
(the non-Cervera terms) are of order $\epsilon^{2} \simeq 10^{-3}$, and 
hence the large gap between the exact and the Cervera {\it et al.} formula, 
in particular the one at $L=4000$ km, must arise as a result of some enhancement. 
It is caused by the factor $1/(1 - r_{A})$, as seen from the relevant energies of 
the gap and the features of the other channels and hierarchies; 
It is nothing but the enhancement due to the MSW resonance \cite{MS1,MS2}. 
It is interesting that our formula, though perturbative, can incorporate the 
enhancement effect for baselines up to several thousand km for which 
$\Delta L \simeq 5$, not too far from $\sim \mathcal{O}(1)$. 
For baseline of order $L \sim 10000$ km the resonance enhancement becomes 
more significant and the difference between the exact and our formula (as well as the Cervera {\it et al.}'s) blows up.
Of course, it is outside the region of validity of the perturbative treatment, and 
one need to sum up $1/(1 - r_{A})$ effect. It is beyond the scope of this paper.

\section{$P (\nu_{e} \rightarrow \nu_{\mu})$ to Order $\epsilon^{5/2}$ and Its $\delta$ Dependence} 
\label{order-5/2}

For large $\theta_{13}$ comparable to the Chooz limit, 
it may be meaningful to analyze the order $\epsilon^{5/2}$ terms, though it has 
an extra suppression of $\sqrt{ \epsilon } \simeq 0.2$ compared to the order 
$\mathcal{ O } \left( \epsilon^{2} \right)$ terms. 
It could be particularly relevant for the 
$\nu_{e} \rightarrow \nu_{\mu}$ channel for which an extreme accuracy may be 
reached e.g., by neutrino factory \cite{ISS-nufact}. 
Our interest in the $\epsilon^{5/2}$ terms is primarily due to that they consist 
only of $\delta$-dependent terms, the property enforced by the theorem A.
Fortunately, it is easy to compute the order $\epsilon^{5/2}$ terms in the 
oscillation probability by using the $\tilde{S}$ matrix elements to order 
$\epsilon^{2}$  listed in Appendix~\ref{tilde-S}. 

The results of the order $\epsilon^{5/2}$ terms in 
$P (\nu_{e} \rightarrow \nu_{\mu})$ which is to be added to the 
$\epsilon^{1}$, $\epsilon^{3/2}$, and $\epsilon^{2}$ terms given in 
Sec.~\ref{e-alpha} reads: 
%
%
\begin{eqnarray}
&& P_{e \mu}^{(5/2)} = 
8 J_{r} s^2_{13} \frac{ r_{\Delta} r_{A} }{ (1 - r_{A})^3 } 
\cos \delta 
\sin^2 \frac{ (1 - r_{A}) \Delta L }{ 2 } 
\nonumber \\
&+& 
8 J_{r} \frac{ r_{\Delta} }{ r_{A} (1 - r_{A}) } 
\left[ 
- 2 s^2_{13}  \frac{ r_{A} }{ (1 - r_{A})^2 } 
+ (c^2_{12} - s^2_{12}) \frac{ r_{\Delta} }{ r_{A} } 
+ s^2_{12} \frac{ r_{\Delta} r_{A} }{ 1 - r_{A} } 
\right] 
\nonumber \\
&& \hspace{60mm}
\times 
\cos \left( \delta - \frac{ \Delta L }{ 2 } \right)
\sin \frac{ r_{A} \Delta L }{ 2 } 
\sin \frac{ (1 - r_{A}) \Delta L }{ 2 }  
\nonumber \\
&+& 
8 J_{r} s^2_{13}  \frac{ r_{\Delta} }{ (1 - r_{A})^2 } (\Delta L ) 
\cos \left( \delta - \frac{ \Delta L }{ 2 } \right)
\sin \frac{ r_{A} \Delta L }{ 2 } 
\cos \frac{ (1 - r_{A}) \Delta L }{ 2 }  
\nonumber \\
&-& 
4 J_{r} s^2_{12} \frac{ r_{\Delta}^2 }{ r_{A} (1 - r_{A}) } (\Delta L ) 
\cos \left( \delta - \frac{ r_{A} \Delta L }{ 2 } \right) 
\sin \frac{ r_{A} \Delta L }{ 2 } 
\nonumber \\
&-&
4 J_{r} c^2_{12} \frac{ r_{\Delta}^2 }{ r_{A} (1 - r_{A}) } (\Delta L ) 
\cos \left( \delta - \frac{ (1 + r_{A}) \Delta L }{ 2 } \right)
\sin \frac{ (1 - r_{A}) \Delta L }{ 2 } 
\nonumber \\
&-&
4 J_{r} \frac{ r_{\Delta} }{ r_{A} (1 - r_{A}) } 
\left( s^2_{13} \frac{ r_{A}  }{ 1 - r_{A} } -  s^2_{12} r_{\Delta} \right) (\Delta L ) 
\cos \left( \delta - \frac{ (1 - r_{A}) \Delta L }{ 2 } \right)
\sin \frac{ (1 - r_{A}) \Delta L }{ 2 }. 
\nonumber \\
\label{Pemu-5/2}
\end{eqnarray}
$P_{e \tau}^{(5/2)}$ can be obtained from 
$P_{e \mu}^{(5/2)}$ by the transformation 
$c_{23} \rightarrow - s_{23}$ and $s_{23} \rightarrow c_{23}$.

The distinctive feature of order $\epsilon^{5/2}$ terms in (\ref{Pemu-5/2}), 
as compared to $\epsilon^{3/2}$ terms in (\ref{Pemu-3/2}), is that 
there exist much more profound type of $\delta$ dependence, 
not only correlated (as argument of cosine function) with $\frac{ \Delta L }{ 2 }$ 
but also to 
$\frac{ r_{A} \Delta L }{ 2 }$ and $\frac{ (1 \pm r_{A}) \Delta L }{ 2 }$. 
If measurement is sufficiently accurate to resolve such correlations different from 
the vacuum type, it must merit to achieve higher sensitivity to detect CP violation 
and accurately measure $\delta$. 
However, quantitative analysis to examine this feature to reveal the required 
experimental conditions, such as which energy resolution and how many 
baselines are required etc. is beyond the scope of this paper.

\section{Parameter Determination and Degeneracy}
\label{degeneracy}

The analysis of parameter degeneracy e.g., in \cite{MU-Pdege} can, in principle, 
be repeated with our formula with higher-order corrections of $s_{13}$. 
However, since the probability contains quartic terms of $s_{13}$, one 
has to deal with eighth-order equations of $s_{13}$ to obtain the full degeneracy 
solutions, a formidable task to carry out. 
Fortunately, we know empirically that apparently there is no other solution beside 
the known eight-fold degeneracy for $\theta_{13}$ below the Chooz limit 
\cite{uchinami}, though it may worth further examination. 
Therefore, in this paper we limit ourselves to the known degeneracy solutions and 
estimate corrections to them due to the large-$\theta_{13}$ correction terms.

We first discuss relationship between the determined mixing parameters 
with and without the non-Cervera terms. 
Given the ``observable'', a pair of the oscillation probabilities 
$P_{e \mu} \equiv P (\nu_{e} \rightarrow \nu_{\mu})$ and 
$\bar{P}_{e \mu} \equiv P (\bar{\nu}_{e} \rightarrow \bar{\nu}_{\mu})$ 
at certain neutrino energy $E$, we consider the 
problem of determining the set of parameters $(s \equiv s_{13}, \delta)$. 
If we use the Cervera {\it et al.} formula, the observable are related to mixing 
parameters $(s_{C}, \delta_{C})$ as 
\begin{eqnarray} 
P_{e \mu} - Z &=& 
X_{\pm}s_{C}^2 + 
Y_{\pm} s_{C} \left( 
\cos \delta_{C} \cos \Delta_{31} \pm \sin \delta_{C} \sin \Delta_{31} 
\right), 
\nonumber \\
\bar{P}_{e \mu} - Z &=& 
X_{\mp}s_{C}^2 - 
Y_{\mp} s_{C} \left( 
\cos \delta_{C} \cos \Delta_{31} \mp \sin \delta_{C} \sin \Delta_{31} 
\right), 
\label{CP-conjugate-C}
\end{eqnarray}
where $\Delta_{31} \equiv \vert  \Delta L / 2 \vert$, and 
$X$, $Y$, and $Z$ are the coefficients whose explicit forms may be constructed 
from the formulas given in Sec.~\ref{oscillation-P} (or, see equation (2.7) 
in \cite{MU-Pdege}). 
The $\pm$ signs in this section imply the normal and the inverted mass hierarchies.

When the non-Cervera corrections are taken into account the same observable 
yield slightly different set $(s_{T}, \delta_{T})$ of mixing parameters as 
\begin{eqnarray} 
P_{e \mu} - Z &=& 
X_{\pm}s_{T}^2 + 
Y_{\pm} s_{T} \left( 
\cos \delta_{T} \cos \Delta_{31} \pm \sin \delta_{T} \sin \Delta_{31} 
\right) 
+ P_{NC}, 
\nonumber \\
\bar{P}_{e \mu} - Z &=& 
X_{\mp} s_{T}^2 - 
Y_{\mp} s_{T} \left( 
\cos \delta_{T} \cos \Delta_{31} \mp \sin \delta_{T} \sin \Delta_{31} 
\right) 
+ \bar{P}_{NC}, 
\label{CP-conjugate-T}
\end{eqnarray}
where $P_{NC}$ and $\bar{P}_{NC}$ denote the non-Cervera corrections in 
neutrino and anti-neutrino channels, respectively. 
The point is that the arguments in $P_{NC}$ and $\bar{P}_{NC}$ is in principle 
$(s_{T}, \delta_{T})$ but it can be replaced by $(s_{C}, \delta_{C})$, the 
known quantities, because all the non-Cervera corrections are second order in 
$\epsilon$. It is then easy to compute the difference between $s_{T}$ and $s_{C}$ 
and $\delta$'s to leading orders in $\epsilon$ as 
\begin{eqnarray} 
\xi_{\pm} \equiv s_{T} - s_{C} =  
-\frac{ 
Y_{\mp} \sin \left( \delta_{C} \pm \Delta_{31} \right) P_{NC} + 
Y_{\pm} \sin \left( \delta_{C} \mp \Delta_{31} \right) \bar{P}_{NC} 
} 
{2 s_{C} \left[ 
X_{\pm} Y_{\mp} \sin \left( \delta_{C} \pm \Delta_{31} \right) + 
X_{\mp} Y_{\pm} \sin \left( \delta_{C} \mp \Delta_{31} \right) 
\right] 
\pm Y_{\pm} Y_{\mp} \sin 2 \Delta_{31} }, 
\\
\eta_{\pm} \equiv \delta_{T} - \delta_{C} = 
\frac{ \left[ 2 X_{\mp} s_{C} - Y_{\mp} \cos \left( \delta_{C} \pm \Delta_{31} \right) \right]  P_{NC} - 
\left[ 2 X_{\pm} s_{C} + Y_{\pm} \cos \left( \delta_{C} \mp \Delta_{31} \right) \right]  \bar{P}_{NC} } 
{2 s_{C}^2 \left[ 
X_{\pm} Y_{\mp} \sin \left( \delta_{C} \pm \Delta_{31} \right) + 
X_{\mp} Y_{\pm} \sin \left( \delta_{C} \mp \Delta_{31} \right) 
\right] 
\pm s_{C} Y_{\pm} Y_{\mp} \sin 2 \Delta_{31} }. 
\label{param-shift}
\end{eqnarray}
It should be noticed that, given $X_{\pm} \sim \mathcal{O} (1)$, $Y_{\pm} \sim \mathcal{O} (\epsilon)$, and $P_{NC} \sim \bar{P}_{NC} \sim \mathcal{O} (\epsilon^2)$, 
$\xi$ and $\eta$ are of order $\epsilon^{3/2}$ and $\epsilon^{1/2}$, respectively. 
They are small compared to $s_{T}$ (or $s_{C}$) and $\delta$ which are of order 
$\epsilon^{1/2}$ and $\epsilon^{0}$, respectively, justifying our perturbative treatment. 
Therefore, inclusion of  the non-Cervera corrections does not produce sizable 
difference in the measured parameters. It is expected because of the smallness 
$\sim \epsilon^{2}$ of the non-Cervera corrections.

Now let us discuss the parameter degeneracy. 
Since $s_{T} - s_{C}$ and $\delta_{T} - \delta_{C}$ are small, the degeneracy 
solutions obtained with the Cervera {\it et al.} formula must give good approximations 
to the ones obtained with our second order formulas. 
Therefore, we start from them; 
Suppose that all the degeneracy solutions 
$(s_{C i}, \delta_{C i})$ ($i=II, III, ... VIII$, reserving $i=I$ for the true solution) 
are obtained by using  the Cervera {\it et al.} formula. 
They are formally given by 
\begin{eqnarray} 
s_{C i} &=& f_{i} ( s_{C 1}, \delta_{C 1} ), 
\nonumber \\ 
\delta_{C i} &=& g_{i} ( s_{C 1}, \delta_{C 1} ).
\label{Dsolution-C}
\end{eqnarray}
The explicit expressions of the functions $f$ and $g$ are given in \cite{MU-Pdege}. 
Here, we take, as a concrete example, the case of the sign-$\Delta m^2$ degeneracy 
whose solutions are labeled as $III$ or $IV$.  
We mention how to extend our analysis to other types of degeneracies. 
Then, $s_{T} - s_{C}$, the difference between $s_{13}$ obtained with our and the 
Cervera {\it et al.} formula, is given for the true $I$ and the sign-$\Delta m^2$ 
clone solution $III$ as 
\begin{eqnarray} 
s_{T III} - s_{C III}  &=&  \xi_{\mp} ( s_{C III}, \delta_{C III} )
\nonumber \\ 
\delta_{T III} - \delta_{C III}  &=&  \eta_{\mp} ( s_{C III}, \delta_{C III} )
\label{Dsolution-T}
\end{eqnarray}
Since $s_{C III}$ and $\delta_{C III}$ are given as functions of $s_{C I}$ and 
$\delta_{C I}$, (\ref{Dsolution-T}) with (\ref{Dsolution-C}) parametrically solve 
$s_{T III}$ and $\delta_{T III}$ as functions of $s_{C I}$ and $\delta_{C I}$. 
For the intrinsic degeneracy solutions there is no need to flip the degeneracy sign 
in (\ref{Dsolution-T}). For degeneracy solutions which involve flipping the octant 
of $\theta_{23}$, $X$ and $Z$ in $\xi$ and $\eta$ must be replaced by the ones 
with octant flip, 
$X_{true} \rightarrow X_{false}= \cot^2 \theta_{23} X_{true}$ and 
$Z_{true} \rightarrow Z_{false}= \tan^2 \theta_{23} Z_{true}$, 
as is done in \cite{MU-Pdege}. 
Thus, all the degeneracy solutions obtained with use of our second order formula 
can be obtained by using the ones with the Cervera {\it et al.} formula as an 
intermediate step. 
The difference between the two is small for $s_{13}$, order $\sim \epsilon^{3/2}$, 
and somewhat larger, $\sim \epsilon^{1/2}$ for $\delta$.

\section{Including the Non-Standard Interactions}
\label{NSI}

The effects of possible nonstandard interactions (NSI) that would affect neutrino 
propagation in matter are usually described by adding the following additional 
term in the tilde basis Hamiltonian 
\begin{eqnarray} 
\tilde{H}_{NSI} = \Delta r_{A}  
\left[
\begin{array}{ccc}
\tilde{\varepsilon}_{e e} & \tilde{\varepsilon}_{e \mu} & \tilde{\varepsilon}_{e \tau} \\
\tilde{\varepsilon}_{e \mu}^* & \tilde{\varepsilon}_{\mu \mu} & \tilde{\varepsilon}_{\mu \tau} \\
\tilde{\varepsilon}_{e \tau}^* & \tilde{\varepsilon}_{\mu \tau}^* & \tilde{\varepsilon}_{\tau \tau} 
\end{array}
\right]. 
\label{H-NSI}
\end{eqnarray}
The relationship between the $\tilde{\varepsilon}_{\alpha \beta}$ and 
the $\varepsilon_{\alpha \beta}$ ($\alpha, \beta = e, \mu, \tau$) parameters in the flavor basis 
is defined in (\ref{tilde-hamiltonian}).
We assume, following \cite{NSI-perturbation}, the NSI elements 
$\varepsilon_{\alpha \beta}$ (and hence $\tilde{\varepsilon}_{\alpha \beta}$) 
are all of the order of $\epsilon$. It is a legitimate assumption because NSI 
comes from higher dimensional operators (dimension six or higher) which 
receives suppression of at least $(M_{W} / M_{NP})^2 \sim 10^{-2}$ with new 
physics scale $M_{NP}$. It should be mentioned, however, that the current 
bounds on the NSI parameters are quite loose, $\lsim 0.1-1$ \cite{NSI-bounds}.

At small $\theta_{13} \sim \epsilon$ for which the $\epsilon$ perturbation theory 
is applicable it was shown in \cite{NSI-perturbation} that inclusion of NSI elements 
into the second-order formula can be done just by 
doing replacement (with slight change in the notations from \cite{NSI-perturbation}) 
\begin{eqnarray} 
r_{\Delta} c_{12} s_{12} &\rightarrow& 
r_{\Delta} c_{12} s_{12}  + r_{A} \tilde{\varepsilon}_{e \mu} 
\equiv \Xi,  
\nonumber \\
s_{13} e^{- i \delta}  &\rightarrow& 
s_{13} e^{- i \delta} + r_{A} \tilde{\varepsilon}_{e \tau} 
\equiv \Theta, 
\label{replacement}
\end{eqnarray}
and nothing else, where 
$ \tilde{\varepsilon}_{e \mu} = ( c_{23} \varepsilon_{e \mu} - s_{23} \varepsilon_{e \tau} )$ 
and 
$ \tilde{\varepsilon}_{e \tau} = (s_{23} \varepsilon_{e \mu} + c_{23}  \varepsilon_{e \tau} )$.
The $\nu_{e}$-related oscillation probability is independent of 
$\varepsilon_{\alpha \beta}$ in the $\mu- \tau$ sector as well as $\varepsilon_{e e}$. 
Then, it is an interesting question to ask how the higher order corrections 
of $s_{13}$ fit in into this picture. The second-order formula for 
$P(\nu _e \rightarrow \nu _\mu)$ with NSI in our 
$\sqrt{\epsilon}$ perturbation theory can be obtained via a straightforward computation. 
The result is given by 
$P(\nu _e \rightarrow \nu _\mu)_{NSI} = P(\nu _e \rightarrow \nu _\mu)_{NSI-C} + 
P(\nu _e \rightarrow \nu _\mu)_{NSI-NC} $ where
%
\begin{eqnarray}
&& P(\nu _e \rightarrow \nu _\mu)_{NSI-C} 
\nonumber \\
&=& 
4 \biggl | c_{23} \Xi 
\left( \frac{1}{ r_{A} } \right)
\sin \frac{ r_{A} \Delta L }{2} 
+ s_{23} e^{ i \frac{\Delta}{2} }
\Theta 
\left( \frac{ 1 }{ 1 - r_{A} } \right)
\sin \frac{ (1 - r_{A})  \Delta L }{2}  
\biggr |^2,  
\label{Pemu-NSI-Cervera}
\end{eqnarray}
%
\begin{eqnarray}
&& P(\nu _e \rightarrow \nu _\mu)_{NSI-NC} 
= 4 s^2_{23} s^2_{13}  \left[ 
s^2_{13} \frac{ (1 + r_{A})^2 }{ (1 - r_{A})^4 } 
- 2 s^2_{12} \frac{ r_{\Delta} r_{A} }{ (1 - r_{A})^3 } 
\right] 
\sin^2 \frac{ (1 - r_{A}) \Delta L }{ 2 } 
\nonumber \\
&+& 
8 s^2_{13} 
\frac{ r_{A}  }{ (1 - r_{A})^2 } 
\left[ 
s^2_{23} 
\left( \tilde{\varepsilon} _{e e } - \tilde{\varepsilon} _{\tau \tau } \right) 
\frac{ 1 }{ 1 - r_{A} } 
+ c_{23} s_{23} \vert \tilde{\varepsilon} _{\mu \tau } \vert \cos \tilde{\phi}_{\mu \tau}
\right] 
\sin^2 \frac{ (1 - r_{A}) \Delta L }{ 2 } 
\nonumber \\
&+& 
2 s^2_{23} s^2_{13} 
\frac{ 1 }{ (1 - r_{A})^2 } 
\left[ 
2 s^2_{13} \frac{ r_{A} }{ 1 - r_{A} } 
- s^2_{12}   r_{\Delta} 
- \left( \tilde{\varepsilon} _{e e } - \tilde{\varepsilon} _{\tau \tau } \right)  r_{A}  
\right] 
(\Delta L) \sin (1 - r_{A}) \Delta L 
\nonumber \\
&-& 
8 c_{23} s_{23} s^2_{13} \vert \tilde{\varepsilon} _{\mu \tau } \vert 
\frac{ 1 }{ 1 - r_{A} } 
\cos \left( \tilde{\phi}_{\mu \tau} + \frac{ \Delta L }{2} \right) 
\sin \frac{ r_{A} \Delta L }{2}  \sin \frac{ (1 - r_{A})  \Delta L }{2}, 
\label{Pemu-NSI-non-Cervera}
\end{eqnarray}
%
where the NSI phase $\tilde{\phi}_{\alpha \beta}$ is defined by 
$ \tilde{\varepsilon} _{\alpha \beta} = \vert \tilde{\varepsilon} _{\alpha \beta} \vert e^{i \tilde{\phi}_{\alpha \beta} }$.
Of course, $P(\nu _e \rightarrow \nu _\tau)_{NSI}$ can be obtained by the transformation 
$c_{23} \rightarrow - s_{23}$ and $s_{23} \rightarrow c_{23}$.

The Cervera term (\ref{Pemu-NSI-Cervera}) contains the terms of 
order $\epsilon^{1}$, $\epsilon^{3/2}$, and $\epsilon^{2}$. 
There exist two NSI dependent terms in $P(\nu _e \rightarrow \nu _\mu)_{NSI-C}$ 
in (\ref{Pemu-NSI-Cervera}) of order $\epsilon^{3/2}$ which are proportional to 
$s_{13} \vert \tilde{\varepsilon} _{e \tau } \vert \cos (\tilde{\phi}_{e \tau} + \delta)$ 
or 
$s_{13} \vert \tilde{\varepsilon} _{e \mu } \vert \cos (\tilde{\phi}_{e \mu} + \delta - \frac{ \Delta L }{2} )$. 
%
Therefore, there is a confusion among the three phases $\delta$, $\tilde{\phi}_{e \mu}$, 
and $\tilde{\phi}_{e \tau}$ which is known to exist in the small $\theta$ regime 
\cite{NSI-2phase,NSI-Madrid}, but here with an amplified magnitude for 
large $\theta_{13} \sim \sqrt{ \epsilon }$.

Notice that the NSI elements in the $\nu_{\mu} - \nu_{\tau}$ sector as well as the 
diagonal ones are absent in the second-order formula of $\nu_{e}$-related probabilities obtained with the $\epsilon$ perturbation theory \cite{NSI-perturbation}. 
The decoupling is supported for small $\theta_{13}$ by the actual data analysis 
in which all the NSI elements are taken into account at the same time \cite{NSI-Madrid}. 
However, the formula (\ref{Pemu-NSI-non-Cervera}) tells us that it is no more true 
at order $\epsilon^2$ for large $\theta_{13}$ of the order of the 
Chooz limit. 
It will require reconsideration of how to determine the NSI parameters 
simultaneously with the $\nu$-mass enriched standard model parameters, 
the problem first addressed in \cite{NSI-perturbation} whose elaboration is 
beyond the scope of this paper.

Though the decoupling between the $\nu_{e}$-related NSI elements and the ones 
in the $\nu_{\mu} - \nu_{\tau}$ sector does not survive, a remnant still remains.
The non-Cervera terms, $P(\nu _e \rightarrow \nu _\mu)_{NSI-NC}$ in 
(\ref{Pemu-NSI-non-Cervera}), consist only of order $\epsilon^2$ terms 
despite that the general theorems discussed in Sec.~\ref{G-feature} do not 
appear to apply to guarantee this property for the system with NSI. 
It is the $\mathcal{O} (\epsilon^2)$ nature of the non-Cervera terms that leads 
to a remnant of decoupling, absence of $\nu_{e}$-related NSI elements in the 
non-Cervera terms;\footnote{
It can be argued that this property holds without doing any computations, 
as was done in the ArXiv version of this paper, arXiv:1103.4387v1 [hep-ph]. 
}
%
The terms induced by the substitution (\ref{replacement}) produce only higher-order terms of order $\sim \epsilon^{5/2}$.

\section{Conclusion}
\label{conclusion}

In this paper, we have constructed a perturbative framework dubbed as 
``$\sqrt{ \epsilon }$ perturbation theory'' to systematically compute higher-order 
corrections of $s_{13}$ assuming that it is large, of the order of the Chooz limit. 
Despite a natural expectation that it could produce sizable corrections to the 
Cervera {\it et al.} formula, we have proven that they must be small, of the order of 
$\epsilon^{2} \simeq 10^{-3}$, where $\epsilon \equiv \Delta m^2_{21} / \Delta m^2_{31}$. 
Nonetheless, we have observed that the correction terms nicely fill the gap between 
the exact  oscillation probability and the Cervera {\it et al.} formula at baseline 
of several thousand km in a limited range of energy where they have enhancement 
due to the resonance effect.

Possible large value of $\theta_{13}$ may allow detection of the $\delta$ dependent 
terms of $\sim \mathcal{O} \left( \epsilon^{5/2} \right)$ in future super-precision 
measurement because they are small only by a factor of $\sim 5$ compared to the 
$\mathcal{O} \left( \epsilon^{2} \right)$ terms. Therefore, 
we have computed the terms and found that they have different $\delta$ dependence 
from the vacuum effect, in correlation to the matter dependent effects. 
However, it remains to be seen if these complicated and coexisting correlations can 
be resolved by actual experimental settings.

Several characteristic features of the computed results prompted us to think about 
some general features of the $\delta$ dependence of the oscillation probability. 
It resulted into the two general theorems stated and proved in 
Sec.~\ref{G-feature} and Appendix~\ref{proof-B}. 
One of them (Theorem A) allows to understand why half-integral order terms in 
$\epsilon$ are always accompanied with cosine or sine $\delta$. 
While the other one (Theorem B) illuminates that the $\delta$ dependence in 
the oscillation probability, both cosine and sine, must be suppressed at the 
small mixing angle or $\Delta m^2_{21} / \Delta m^2_{31} \rightarrow 0$ limits. 
Both of them cooperate to illuminate some characteristic features of the 
perturbatively computed oscillation probabilities

We have investigated effects of the non-Cervera correction terms on 
parameter determination and the degeneracy. 
Because the correction terms are small, their effect must be small. 
Nonetheless, the explicitly computed corrections to the degeneracy solutions 
obtained with use of the Cervera {\it et al.} formula may be of use when they 
are implemented into the analysis codes such as the ones described in 
\cite{Globes,M-cube}.\footnote{
We thank Patrick Huber for communications for feasibility of  implementing the 
analytic degeneracy solutions to achieve fast search for the degenerate minima, 
the possibility raised in \cite{MU-Pdege}. 
}

Finally, we gave a derivation of the second-order formula of the 
$\nu_{e}$-related appearance probabilities with large $\theta_{13}$ 
corrections in systems with NSI effects in propagation. 
The result of the Cervera terms is shown to be identical with the one obtained by 
the replacement to the generalized variables (\ref{replacement}), producing 
enhanced NSI dependent terms of $\mathcal{O} \left( \epsilon^{3/2} \right)$. 
The decoupling of the NSI elements in the $\nu_{\mu} - \nu_{\tau}$ sector 
from the $\nu_{e}$-related probabilities is invalidated by the non-Cervera 
type corrections at order $\epsilon^2$, which calls for reconsideration of the 
strategy for parameter determination.

\appendix

\section{Theorem B for Suppression of CP Violation for Vanishing Mixing Angle}
\label{proof-B}

Here, we attempt to prove the theorem B. 
We first describe our general strategy. 
We introduce a diagonal phase transformation matrix 
$T = \text{diag} (e^{i \alpha}, e^{i \beta},  e^{i \gamma})$, and define a 
``hat basis'' as $\hat{\nu} = T \nu$ with the Hamiltonian 
$\hat{H} \equiv T H T^{\dagger}$. 
We choose  $\alpha$, $\beta$, and $\gamma$ such that $\hat{H}$ is free from 
any phases including $\delta$. 
Since $T$ is a diagonal phase transformation, it merely redefines phase of the 
flavor basis wave function, and hence it does not affect the probability. 
Therefore, if such choice of the phases is shown to be possible it implies that 
there is no $\delta$ dependence in the oscillation probability.

We illustrate the proof by explicitly treating the case of $\theta_{23} = 0$ case, 
because the situation is exactly the same in other cases. 
With the choice $T = \text{diag} (1, 1, e^{- i \delta})$, the hat basis Hamiltonian 
can be written as $\hat{H} = \hat{H}_{vac} + \text{diag} (a/2E, 0, 0)$, where 
\begin{eqnarray} 
\hat{H}_{vac} &=& 
  \left[ 
  \begin{array}{ccc}
  \Delta_{21} s_{12}^2 c_{13}^2 + \Delta_{31} s_{13}^2 &  \Delta_{21} c_{12} s_{12} c_{13} &
  - \Delta_{21} s_{12}^2 s_{13} c_{13} + \Delta_{31} c_{13} s_{13} \\
   \Delta_{21} c_{12} s_{12} c_{13} &   \Delta_{21} c_{12}^2 &
  - \Delta_{21} c_{12} s_{12} s_{13} \\
  - \Delta_{21} s_{12}^2 s_{13} c_{13} + \Delta_{31} c_{13} s_{13} &  - \Delta_{21} c_{12} s_{12} s_{13}  &  \Delta_{21} s_{12}^2 s_{13}^2 + \Delta_{31} c_{13}^2
  \end{array}
  \right] 
 \label{hatH-23}
\end{eqnarray}
showing explicitly that $\delta$-dependence disappear from the Hamiltonian by 
the $T$ transformation. Notice that here we have used the notation 
$\Delta_{ij} \equiv \Delta m^2_{ij} / 2E$ ($ij=21, 31$), whose latter is different 
from the one in Sec.~\ref{degeneracy}. (We hope that no confusion arises.) 
Therefore, $\delta$ goes away from the probability 
in the vanishing $\theta_{23}$ limit.
We can repeat the similar exercise for other vanishing limits of 
$s_{12}$ with $T$ matrix $T = \text{diag} (e^{i \delta}, 1, 1)$. 
It is trivial to observe that there is no $\delta$ dependence if $\theta_{13}=0$.
Therefore, we have shown that $\delta$ dependence, 
both $\cos \delta$ and $\sin \delta$, disappears from the oscillation probability 
at the vanishing limit of one of the mixing angles for arbitrary matter density profiles.

To really prove the theorem B we have to show that the $\delta$-dependence 
goes away in the large mixing angle limit $\theta_{ij} \rightarrow \pi/2$ 
($ij = 12$ and 23). 
One can show that the similar method works for $c_{12} \rightarrow 0$ and 
$c_{23} \rightarrow 0$ with $T = \text{diag} (e^{i \delta}, 1, 1)$ and 
$T = \text{diag} (1, e^{- i \delta}, 1)$, respectively. 
Assuming that the oscillation probabilities in all channels can be expanded into 
power series of $s_{ij}$ and $c_{ij}$ the features stated above prove the theorem B 
for arbitrary matter density profiles.

It is interesting to observe that the similar method fails for the limit 
$c_{13} \rightarrow 0$. It is perfectly consistent with the fact the factor $c_{13}$ 
is missing in the coefficient of $\cos \delta$ term in the oscillation probabilities in 
the $\nu_{\mu}-\nu_{\tau}$ sector \cite{KTY2}. 
However, since the suppression factor of the $\delta$-dependent terms in the 
$\nu_{e}$-related channels for constant matter density derived in the same 
reference \cite{KTY2} {\em does} contain $c^2_{13}$, it is likely that 
the theorem B can be generalized to the one with suppression factor $J$ instead 
of $J_{r}$ in these channels, a conjecture.

\section{$\tilde{S}$ matrix elements to second order}
\label{tilde-S}

The $\tilde{S}$ matrix elements can be computed by using (\ref{Smatrix-tilde2}) 
and are written as sums over the terms of order 
$\epsilon^{\frac{1}{2}}$, $\epsilon^{1}$, $\epsilon^{\frac{3}{2}}$, and $\epsilon^{2}$. 
The T-conjugate elements can be obtained by 
$\tilde{S}_{\beta \alpha} (\delta) = \tilde{S}_{\alpha \beta} (- \delta)$. 
Notice that all the elements that are not shown below, except for the T-conjugate 
ones, are vanishing. 
\begin{eqnarray}
\tilde{S}^{(1/2)}_{e \tau}  =
s_{13} e^{ - i \delta} \frac{ 1 }{ 1 - r_{A} } 
\left( e^{- i \Delta x } - e^{- i r_{A} \Delta x } \right) 
\label{tilde-S-1/2}
\end{eqnarray}
%
\begin{eqnarray}
\tilde{S}^{(1)}_{ee} &=& 
 \left( s^2_{13} \frac{ r_{A} }{ 1 - r_{A} } - s^2_{12} r_{\Delta} \right) ( i \Delta x )  e^{- i r_{A} \Delta x}
+ s^2_{13} \frac{ 1 }{ (1 - r_{A})^2 } 
\left( e^{- i \Delta x } - e^{- i r_{A} \Delta x } \right) 
\nonumber \\
\tilde{S}^{(1)}_{e \mu} &=& 
- c_{12} s_{12} \left( \frac{ r_{\Delta} }{ r_{A} } \right) 
\left( 1 -  e^{- i r_{A} \Delta x} \right) 
\label{tilde-S-1-e}
\end{eqnarray}
%
\begin{eqnarray}
\tilde{S}^{(1)}_{\mu \mu} &=& 
- c^2_{12} r_{\Delta} ( i \Delta x ) 
\nonumber \\
\tilde{S}^{(1)}_{\tau \tau} &=& 
- s^2_{13} \left( \frac{ r_{A} }{ 1 - r_{A} } \right) ( i \Delta x )  e^{- i \Delta x}
- s^2_{13} \frac{ 1 }{ (1 - r_{A})^2 } 
\left( e^{- i \Delta x } - e^{- i r_{A} \Delta x } \right) 
\label{tilde-S-1-mutau}
\end{eqnarray}
%
\begin{eqnarray} 
\tilde{S}^{(3/2)}_{e \tau} &=& 
- s^3_{13} e^{- i \delta} \frac{  (1 + r_{A})^2 }{ 2 (1 - r_{A})^3 } 
\left( e^{- i \Delta x } - e^{- i r_{A} \Delta x } \right) 
\nonumber \\
&-& 
s^3_{13} e^{- i \delta} \frac{  r_{A} }{ (1 - r_{A})^2 } (i \Delta x) \left( e^{- i \Delta x } + e^{- i r_{A} \Delta x } \right) 
\nonumber \\
&+& 
s^2_{12} s_{13} e^{- i \delta} 
\frac{  r_{\Delta}  }{ 1 - r_{A} } 
\left[ 
\frac{ r_{A} }{ 1 - r_{A} }\left( e^{- i \Delta x } - e^{- i r_{A} \Delta x } \right) + 
\left( i \Delta x \right) e^{- i r_{A} \Delta x } 
\right] 
\nonumber \\
\tilde{S}^{(3/2)}_{\mu \tau} &=& 
- c_{12} s_{12} s_{13} e^{- i \delta} 
\frac{  r_{\Delta}  }{ 1 - r_{A} } 
\left[ 
r_{A}  \left( 1 - e^{- i \Delta x } \right) - 
\frac{  1 }{ r_{A} } \left( 1 - e^{- i r_{A} \Delta x } \right) 
\right] 
\label{tilde-S-3/2}
\end{eqnarray}
%

\begin{eqnarray} 
\tilde{S}^{(2)}_{ee} &=& 
\left( s^2_{12} r_{\Delta} - s^2_{13} \frac{ r_{A} }{ 1 - r_{A} }  \right)^2 
\frac{ ( i \Delta x )^2 }{ 2 } e^{- i r_{A} \Delta x} 
- s^4_{13} \frac{ r_{A} }{ (1 - r_{A})^3 } 
( i \Delta x ) e^{- i \Delta x} 
\nonumber \\
&-& 
\left[ 
 c^2_{12} s^2_{12} \left( \frac{ r_{\Delta}^2 }{ r_{A} } \right) 
- s^2_{12} s^2_{13} r_{\Delta} \frac{ 1 + r_{A}^2 }{ (1 - r_{A})^2 } 
+ s^4_{13} \frac{ r_{A} (1 + r_{A}) }{ (1 - r_{A})^3 } 
\right] 
( i \Delta x ) e^{- i r_{A} \Delta x} 
\nonumber \\
&+& 
\left[
2 s^2_{12}  s^2_{13} r_{\Delta} \frac{ r_{A} }{ (1 - r_{A})^3 } 
-  s^4_{13} \frac{ r_{A} ( 2 + r_{A} ) }{ (1 - r_{A})^4 }
\right] 
\left( e^{- i \Delta x} - e^{- i r_{A} \Delta x}  \right) 
\nonumber \\
&+&
c^2_{12} s^2_{12} \left( \frac{ r_{\Delta} }{ r_{A} } \right)^2 
\left( 1 -  e^{- i r_{A} \Delta x} \right) 
\label{tilde-See-2}
\end{eqnarray}

\begin{eqnarray} 
\tilde{S}^{(2)}_{e \mu} &=& 
c^3_{12} s_{12} 
\left( \frac{ r_{\Delta}^2 }{ r_{A} } \right) ( i \Delta x ) 
- c_{12} s_{12} 
\left( \frac{ r_{\Delta} }{ r_{A} } \right) 
\left[
s^2_{12} r_{\Delta} - s^2_{13}  \frac{ r_{A} }{ 1 - r_{A} } 
 \right] 
( i \Delta x ) e^{- i r_{A} \Delta x} 
\nonumber \\
&-& 
c_{12} s_{12}  \left( \frac{ r_{\Delta} }{ r_{A} } \right) 
\left[ 
\frac{1}{2} s^2_{13}
+ \left( c^2_{12} -  s^2_{12} \right) 
 \left( \frac{ r_{\Delta} }{ r_{A} } \right) 
\right] 
\left( 1 -  e^{- i r_{A} \Delta x} \right) 
\nonumber \\
&+& c_{12} s_{12} s^2_{13} \frac{ r_{\Delta} r_{A} }{ (1 - r_{A})^2 } 
\left( e^{- i \Delta x} - e^{- i r_{A} \Delta x}  \right) 
\end{eqnarray}
%

\begin{eqnarray} 
%
\tilde{S}^{(2)}_{\mu \mu} &=& 
c^4_{12} r_{\Delta}^2 \frac{ ( i \Delta x )^2 }{ 2 } 
+ c^2_{12} s^2_{12} \left( \frac{ r_{\Delta}^2 }{  r_{A}  } \right) ( i \Delta x ) 
- c^2_{12} s^2_{12} \left( \frac{ r_{\Delta} }{ r_{A} } \right)^2 
\left( 1 -  e^{- i r_{A} \Delta x} \right) 
\label{tilde-Smumu-2}
\end{eqnarray}

\begin{eqnarray} 
\tilde{S}^{(2)}_{\tau \tau} &=& 
s^4_{13} \left( \frac{ r_{A} }{ 1 - r_{A} } \right)^2
\frac{ ( i \Delta x )^2 }{ 2 } e^{- i \Delta x} 
\nonumber \\
&+& 
\left[ 
s^4_{13} \frac{ r_{A} (1 + r_{A}) }{ (1 - r_{A})^3 } 
- s^2_{12} s^2_{13} r_{\Delta} 
\left( \frac{ r_{A} }{ 1 - r_{A} } \right)^2 
\right] 
( i \Delta x ) e^{- i  \Delta x} 
\nonumber \\
&+& 
\left[
s^4_{13}  \frac{ r_{A} }{ (1 - r_{A})^3 } - s^2_{12} s^2_{13} r_{\Delta} 
\frac{ 1 }{ (1 - r_{A})^2 } 
\right] 
( i \Delta x ) e^{- i r_{A} \Delta x} 
\nonumber \\
&+& 
\left[
s^4_{13} \frac{ r_{A} ( 2 + r_{A} ) }{ (1 - r_{A})^4 }
- 2 s^2_{12}  s^2_{13} r_{\Delta} \frac{ r_{A} }{ (1 - r_{A})^3 } 
\right] 
\left( e^{- i \Delta x} - e^{- i r_{A} \Delta x}  \right) 
\label{tilde-Stautau-2}
\end{eqnarray}

The $S$ matrix elements can be obtained from the $\tilde{S}$ matrix elements 
by 
\begin{eqnarray}
&& S \equiv 
\left[
\begin{array}{ccc}
S_{ee} & S_{e \mu} & S_{e \tau} \\
S_{\mu e} & S_{\mu \mu} & S_{\mu \tau} \\
S_{\tau e} & S_{\tau \mu} & S_{\tau \tau} 
\end{array}
\right] 
\nonumber \\
&=&
\left[
\begin{array}{ccc}
\tilde{ S }_{ee} & c_{23} \tilde{ S }_{e \mu} + s_{23} \tilde{ S }_{e \tau} & - s_{23} \tilde{ S }_{e \mu} + c_{23} \tilde{ S }_{e \tau}  \\
c_{23} \tilde{ S }_{\mu e} + s_{23} \tilde{ S }_{\tau e} & c^2_{23} \tilde{ S }_{\mu \mu} + s^2_{23} \tilde{ S }_{\tau \tau} + c_{23} s_{23} ( \tilde{ S }_{\mu \tau} + \tilde{ S }_{\tau \mu} ) & c^2_{23} \tilde{ S }_{\mu \tau} - s^2_{23} \tilde{ S }_{\tau \mu} + c_{23} s_{23} ( \tilde{ S }_{\tau \tau} - \tilde{ S }_{\mu \mu} ) \\
- s_{23} \tilde{ S }_{\mu e} + c_{23} \tilde{ S }_{\tau e} & c^2_{23} \tilde{ S }_{\tau \mu } - s^2_{23} \tilde{ S }_{ \mu \tau} + c_{23} s_{23} ( \tilde{ S }_{\tau \tau} - \tilde{ S }_{\mu \mu} ) & s^2_{23} \tilde{ S }_{\mu \mu} + c^2_{23} \tilde{ S }_{\tau \tau} - c_{23} s_{23} ( \tilde{ S }_{\mu \tau} + \tilde{ S }_{\tau \mu} ) 
\end{array}
\right]. 
\nonumber \\
\label{SbyS-tilde}
\end{eqnarray}

\section{Second Order Expressions of Oscillation Probabilities}
\label{probability-app}

In this Appendix~\ref{probability-app} we present some remaining expressions of 
the oscillation probabilities. 
In the $\nu_e \rightarrow \nu_\tau$ channel, 
$P_{e \tau}^{(i)}$ ($i=1, 3/2, 2$) are given in each order by 
%
\begin{eqnarray}
P_{e \tau}^{(1)} &=& 
4 c^2_{23} s^2_{13} \frac{ 1 }{ (1 - r_{A})^2 } 
\sin^2 \frac{ (1 - r_{A}) \Delta L }{ 2 }, 
\label{Petau-1}
\end{eqnarray}
%
\begin{eqnarray}
P_{e \tau}^{(3/2)} &=& 
- 8 J_{r} \frac{ r_{\Delta} }{ r_{A} (1 - r_{A}) }
\cos \left( \delta - \frac{ \Delta L }{ 2 } \right)
\sin \frac{ r_{A} \Delta L }{ 2 } 
\sin \frac{ (1 - r_{A}) \Delta L }{ 2 }, 
\label{Petau-3/2}
\end{eqnarray}
%
\begin{eqnarray}
P_{e \tau}^{(2)} &=& 
4 s^2_{23} c^2_{12} s^2_{12} 
\left( \frac{ r_{\Delta} }{ r_{A} } \right)^2 
\sin^2 \frac{ r_{A} \Delta L }{ 2 } 
\nonumber \\
&-& 
4 c^2_{23} \left[ 
s^4_{13} \frac{ (1 + r_{A})^2 }{ (1 - r_{A})^4 } 
- 2 s^2_{12} s^2_{13} \frac{ r_{\Delta} r_{A} }{ (1 - r_{A})^3 } 
\right] 
\sin^2 \frac{ (1 - r_{A}) \Delta L }{ 2 }. 
\nonumber \\
&+& 
2 c^2_{23} \left[ 
2 s^4_{13} \frac{ r_{A} }{ (1 - r_{A})^3 } 
- s^2_{12} s^2_{13} \frac{ r_{\Delta} }{ (1 - r_{A})^2 } 
\right] 
(\Delta L ) \sin (1 - r_{A}) \Delta L. 
\label{Petau-2}
\end{eqnarray}
While in the $\nu_\mu \rightarrow \nu_\tau$ channel, 
$P_{\mu \tau}^{(i)}$ ($i=0, 1, 3/2, 2$) are given by
%
%
\begin{eqnarray}
P_{\mu \tau}^{(0)} &=& 
4 c^2_{23}  s^2_{23} \sin^2 \left( \frac{ \Delta L }{ 2 } \right), 
\label{Pmutau-0}
\end{eqnarray}
\begin{eqnarray}
P_{\mu \tau}^{(1)} &=& 
2 c^2_{23}  s^2_{23} 
\left[ 
s^2_{13} \frac{ r_{A} }{ 1 - r_{A} } - c^2_{12} r_{\Delta} 
\right]  
( \Delta L ) \sin \Delta L 
\nonumber \\
&-& 
8 c^2_{23}  s^2_{23} s^2_{13} \frac{ 1 }{ (1 - r_{A})^2 } 
\sin \left( \frac{ \Delta L }{ 2 } \right) 
\cos \left( \frac{ r_{A} \Delta L }{ 2 } \right)
\sin \frac{ (1 - r_{A}) \Delta L }{ 2 }, 
\label{Pmutau-1}
\end{eqnarray}
\begin{eqnarray}
P_{\mu \tau}^{(3/2)} &=& 
8 J_{r} \cos \delta \left( c^2_{23} - s^2_{23} \right) 
\frac{ r_{\Delta}  }{ 1 - r_{A} } 
\nonumber \\
&\times&
\left[ 
r_{A} \sin^2 \left( \frac{ \Delta L }{ 2 } \right) 
- \frac{ 1 }{ r_{A} } 
\sin \left( \frac{ \Delta L }{ 2 } \right) 
\sin \left( \frac{ r_{A} \Delta L }{ 2 } \right)
\cos \frac{ (1 - r_{A}) \Delta L }{ 2 } 
\right] 
\nonumber \\
&+& 
8 J_{r} \sin \delta \frac{ r_{\Delta}  }{ r_{A} (1 - r_{A}) } 
\sin \left( \frac{ \Delta L }{ 2 } \right) 
\sin \left( \frac{ r_{A} \Delta L }{ 2 } \right)
\sin \frac{ (1 - r_{A}) \Delta L }{ 2 }, 
\label{Pmutau-3/2}
\end{eqnarray}
\begin{eqnarray}
&& P_{\mu \tau}^{(2)} = 
c^2_{23}  s^2_{23}  
\left( s^2_{13} \frac{ r_{A} }{ 1 - r_{A} } - c^2_{12} r_{\Delta}  \right)^2 
( \Delta L )^2 \cos \Delta L 
\nonumber \\
&-& 
2 c^2_{23}  s^2_{23} 
\left[
s^4_{13}  \frac{ r_{A} (1 + r_{A}) }{ (1 - r_{A})^3 } 
- (c^2_{12} + s^2_{12} r_{A}^2) s^2_{13} \frac{ r_{\Delta} }{ (1 - r_{A})^2 } 
- c^2_{12} s^2_{12}  \frac{ r_{\Delta}^2 }{ r_{A} }
\right] 
( \Delta L ) \sin \Delta L 
\nonumber \\
&-& 
2 c^2_{23}  s^2_{23} 
\left[ 
s^4_{13}  \frac{ r_{A} }{ (1 - r_{A})^3 } 
+ (c^2_{12} - s^2_{12}) s^2_{13} \frac{ r_{\Delta} }{ (1 - r_{A})^2 } 
\right] 
( \Delta L ) \sin r_{A} \Delta L 
\nonumber \\
&-& 
2 c^2_{23}  s^2_{23} 
\left[ 
2 s^4_{13}  \frac{ r_{A} }{ (1 - r_{A})^3 } 
- s^2_{12} s^2_{13} \frac{ r_{\Delta} }{ (1 - r_{A})^2 } 
\right] 
( \Delta L ) \sin (1 - r_{A}) \Delta L 
\nonumber \\
&+& 
4 c^2_{23}  s^2_{23} 
\left[
s^4_{13} \frac{ r_{A} ( 2 + r_{A} ) }{ (1 - r_{A})^4 }
- 2 s^2_{12}  s^2_{13} \frac{ r_{\Delta} r_{A} }{ (1 - r_{A})^3 } 
- c^2_{12} s^2_{12} \left( \frac{ r_{\Delta} }{ r_{A} } \right)^2 
\right] 
\sin^2 \frac{ \Delta L }{ 2 } 
\nonumber \\
&-& 
4 c^2_{23}  s^2_{23} 
\left[
s^4_{13} \frac{ r_{A} ( 2 + r_{A} ) }{ (1 - r_{A})^4 }
- 2 s^2_{12}  s^2_{13} \frac{ r_{\Delta} r_{A} }{ (1 - r_{A})^3 } 
+ c^2_{12} s^2_{12} \left( \frac{ r_{\Delta} }{ r_{A} } \right)^2 
\right] 
\sin^2 \frac{ r_{A} \Delta L }{ 2 } 
\nonumber \\
&+& 
4 c^2_{23}  s^2_{23} 
\left[
s^4_{13} \frac{ ( 1 + r_{A} )^2 }{ (1 - r_{A})^4 }
- 2 s^2_{12}  s^2_{13} \frac{ r_{\Delta} r_{A} }{ (1 - r_{A})^3 } 
+ c^2_{12} s^2_{12} \left( \frac{ r_{\Delta} }{ r_{A} } \right)^2 
\right] 
\sin^2 \frac{ (1- r_{A}) \Delta L }{ 2 }. 
\label{Pmutau-2}
\end{eqnarray}
%

\begin{acknowledgments}
The authors thank Hiroshi Numokawa and Osamu Yasuda for useful comments.
This manuscript was completed while we were receiving numerous heartfelt 
messages from our friends all over the world since March 11, 2011, 
to which we would like to express our deep gratitudes. 

\end{acknowledgments}

\end{document}